%% file: mrk817_swift.tex
\documentclass[twocolumn]{aastex631}
\usepackage{mathrsfs,apjfonts}

\def\mrk{{Mrk~817}}
\def\swift{{Swift}}
\def\hst{{HST}}

\shorttitle{AGN STORM 2: Mrk 817 with Swift}
\shortauthors{Cackett et al.}

\begin{document}

\title{AGN STORM 2. IV. Swift X-ray and ultraviolet/optical monitoring of Mrk 817}

\correspondingauthor{Edward M. Cackett}
\email{ecackett@wayne.edu}

\author[0000-0002-8294-9281]{Edward M. Cackett}
\affiliation{Department of Physics and Astronomy, Wayne State University, 666 W.\ Hancock St, Detroit, MI, 48201, USA}

\author[0000-0001-9092-8619]{Jonathan Gelbord}
\affiliation{Spectral Sciences Inc., 30 Fourth Ave., Suite 2, Burlington, MA 01803, USA}

\author[0000-0002-3026-0562]{Aaron J.\ Barth}
\affiliation{Department of Physics and Astronomy, 4129 Frederick Reines Hall, University of California, Irvine, CA, 92697-4575, USA}

\author[0000-0003-3242-7052]{Gisella De~Rosa}
\affiliation{Space Telescope Science Institute, 3700 San Martin Drive, Baltimore, MD 21218, USA}

\author[0000-0001-8598-1482]{Rick Edelson} 
\affiliation{Eureka Scientific Inc., 2452 Delmer St. Suite 100, Oakland, CA 94602, USA}

\author[0000-0002-2908-7360]{Michael R.\ Goad}
\affiliation{School of Physics and Astronomy, University of Leicester, University Road, Leicester, LE1 7RH, UK}

\author[0000-0002-0957-7151]{Yasaman Homayouni}
\affiliation{Space Telescope Science Institute, 3700 San Martin Drive, Baltimore, MD 21218, USA}
\affiliation{Department of Astronomy and Astrophysics, The Pennsylvania State University, 525 Davey Laboratory, University Park, PA 16802}
\affiliation{Institute for Gravitation and the Cosmos, The Pennsylvania State University, University Park, PA 16802}

\author[0000-0003-1728-0304]{Keith Horne}
\affiliation{SUPA School of Physics and Astronomy, North Haugh, St.~Andrews, KY16~9SS, Scotland, UK}

\author[0000-0003-0172-0854]{Erin A.\ Kara}
\affiliation{MIT Kavli Institute for Astrophysics and Space Research, Massachusetts Institute of Technology, Cambridge, MA 02139, USA}

\author[0000-0002-2180-8266]{Gerard A.\ Kriss}
\affiliation{Space Telescope Science Institute, 3700 San Martin Drive, Baltimore, MD 21218, USA}

\author[0000-0003-0944-1008]{Kirk T.\ Korista}
\affiliation{Department of Physics, Western Michigan University, 1120 Everett Tower, Kalamazoo, MI 49008-5252, USA}

\author{Hermine Landt}
\affiliation{Centre for Extragalactic Astronomy, Department of Physics, Durham University, South Road, Durham DH1 3LE, UK}

\author[0000-0002-2509-3878]{Rachel Plesha}
\affiliation{Space Telescope Science Institute, 3700 San Martin Drive, Baltimore, MD 21218, USA}

\author[0000-0003-2991-4618]{Nahum Arav}
\affiliation{Department of Physics, Virginia Tech, Blacksburg, VA 24061, USA}

\author[0000-0002-2816-5398]{Misty C.\ Bentz}
\affiliation{Department of Physics and Astronomy, Georgia State University, 25 Park Place, Suite 605, Atlanta, GA 30303, USA}

\author[0000-0001-6301-570X]{Benjamin D. Boizelle}
\affiliation{Department of Physics and Astronomy, N284 ESC, Brigham Young University, Provo, UT, 84602, USA}

\author[0000-0001-9931-8681]{Elena Dalla Bont\`{a}}\altaffiliation{Visiting Fellow at UCLan}
\affiliation{Dipartimento di Fisica e Astronomia ``G.\  Galilei,'' Universit\`{a} di Padova, Vicolo dell'Osservatorio 3, I-35122 Padova, Italy}
\affiliation{INAF-Osservatorio Astronomico di Padova, Vicolo dell'Osservatorio 5 I-35122, Padova, Italy}
\affiliation{Jeremiah Horrocks Institute, University of Central Lancashire, Preston, PR1 2HE, UK}

\author[0000-0002-0964-7500]{Maryam Dehghanian}
\affiliation{Department of Physics, Virginia Tech, Blacksburg, VA 24061, USA}

\author[0000-0002-6460-3682]{Fergus Donnan}
\affiliation{Department of Physics, University of Oxford, Keble Road, Oxford, OX1 3RH, UK}

\author[0000-0002-5830-3544]{Pu Du} 
\affiliation{Key Laboratory for Particle Astrophysics, Institute of High Energy Physics, Chinese Academy of Sciences, 19B Yuquan Road,\\ Beijing 100049, People's Republic of China}

\author[0000-0003-4503-6333]{Gary J.\ Ferland}
\affiliation{Department of Physics and Astronomy, The University of Kentucky, Lexington, KY 40506, USA}

\author[0000-0002-2306-9372]{Carina Fian}
\affiliation{Haifa Research Center for Theoretical Physics and Astrophysics, University of Haifa, Haifa 3498838, Israel}
\affiliation{School of Physics and Astronomy and Wise observatory, Tel Aviv University, Tel Aviv 6997801, Israel}

\author[0000-0003-3460-0103]{Alexei V.\ Filippenko}
\affiliation{Department of Astronomy, University of California, Berkeley, CA 94720-3411, USA}

\author[0000-0002-9280-1184]{Diego H.\ Gonz\'{a}lez Buitrago}
\affiliation{Instituto de Astronom\'{\i}a, Universidad Nacional Aut\'{o}noma de M\'{e}xico, Km 103 Carretera Tijuana-Ensenada, 22860 Ensenada B.C., M\'{e}xico}

\author[0000-0001-9920-6057]{Catherine~J.~Grier}
\affiliation{Department of Astronomy, University of Wisconsin-Madison, Madison, WI 53706, USA} 

\author[0000-0002-1763-5825]{Patrick B.\ Hall}
\affiliation{Department of Physics and Astronomy, York University, Toronto, ON M3J 1P3, Canada}

\author{Chen Hu}
\affiliation{Key Laboratory for Particle Astrophysics, Institute of High Energy Physics, Chinese Academy of Sciences, 19B Yuquan Road,\\ Beijing 100049, People's Republic of China}

\author[0000-0002-1134-4015]{Dragana Ili\'{c}}
\affiliation{University of Belgrade - Faculty of Mathematics, Department of Astronomy, Studentski trg 16, 11000 Belgrade, Serbia}
\affiliation{Humboldt Research Fellow, Hamburger Sternwarte, Universit{\"a}t Hamburg, Gojenbergsweg 112, 21029 Hamburg, Germany}

\author[0000-0001-5540-2822]{Jelle Kaastra}
\affiliation{SRON Netherlands Institute for Space Research, Niels Bohrweg 4, 2333 CA Leiden, The Netherlands}
\affiliation{Leiden Observatory, Leiden University, PO Box 9513, 2300 RA Leiden, The Netherlands}

\author[0000-0002-9925-534X]{Shai Kaspi}
\affiliation{School of Physics and Astronomy and Wise observatory, Tel Aviv University, Tel Aviv 6997801, Israel}

\author[0000-0001-6017-2961]{Christopher S.\ Kochanek}
\affiliation{Department of Astronomy, The Ohio State University, 140 W.\ 18th Ave., Columbus, OH 43210, USA}
\affiliation{Center for Cosmology and AstroParticle Physics, The Ohio State University, 191 West Woodruff Ave., Columbus, OH 43210, USA}

\author[0000-0001-5139-1978]{Andjelka B. Kova{\v c}evi{\'c}}
\affiliation{University of Belgrade - Faculty of Mathematics, Department of Astronomy, Studentski trg 16, 11000 Belgrade, Serbia}
\affiliation{Key Laboratory for Particle Astrophysics, Institute of High Energy Physics, Chinese Academy of Sciences, 19B Yuquan Road,\\ Beijing 100049, People's Republic of China}

\author[0000-0001-8638-3687]{Daniel Kynoch}
\affiliation{Astronomical Institute of the Czech Academy of Sciences, Bo\v{c}n\'{i} II 1401/1a, CZ-14100 Prague, Czechia}

\author[0000-0001-5841-9179]{Yan-Rong Li}
\affiliation{Key Laboratory for Particle Astrophysics, Institute of High Energy Physics, Chinese Academy of Sciences, 19B Yuquan Road,\\ Beijing 100049, People's Republic of China}

\author[0000-0003-1081-2929]{Jacob N. McLane}
\affiliation{Department of Physics and Astronomy, University of Wyoming, Laramie, WY 82071, USA}

\author[0000-0002-4992-4664]{Missagh Mehdipour}
\affiliation{Space Telescope Science Institute, 3700 San Martin Drive, Baltimore, MD 21218, USA}

\author[0000-0001-8475-8027]{Jake A. Miller}
\affiliation{Department of Physics and Astronomy, Wayne State University, 666 W.\ Hancock St, Detroit, MI, 48201, USA}

\author[0000-0001-5639-5484]{John Montano}
\affiliation{Department of Physics and Astronomy, 4129 Frederick Reines Hall, University of California, Irvine, CA, 92697-4575, USA}

\author[0000-0002-6766-0260]{Hagai Netzer}
\affiliation{School of Physics and Astronomy and Wise observatory, Tel Aviv University, Tel Aviv 6997801, Israel}

\author{Christos Panagiotou}
\affiliation{MIT Kavli Institute for Astrophysics and Space Research, Massachusetts Institute of Technology, Cambridge, MA 02139, USA}

\author[0000-0003-1183-1574]{Ethan Partington}
\affiliation{Department of Physics and Astronomy, Wayne State University, 666 W.\ Hancock St, Detroit, MI, 48201, USA}

\author[0000-0003-2398-7664]{Luka \v{C}.\ Popovi\'{c}}
\affiliation{Astronomical Observatory, Volgina 7, 11060 Belgrade, Serbia}
\affiliation{University of Belgrade - Faculty of Mathematics, Department of Astronomy, Studentski trg 16, 11000 Belgrade, Serbia}
\affiliation{Key Laboratory for Particle Astrophysics, Institute of High Energy Physics, Chinese Academy of Sciences, 19B Yuquan Road,\\ Beijing 100049, People's Republic of China}

\author[0000-0002-6336-5125]{Daniel Proga}
\affiliation{Department of Physics \& Astronomy, University of Nevada, Las Vegas, 4505 S.\ Maryland Pkwy, Las Vegas, NV, 89154-4002, USA}


\author[0000-0002-5359-9497]{Daniele Rogantini}
\affiliation{MIT Kavli Institute for Astrophysics and Space Research, Massachusetts Institute of Technology, Cambridge, MA 02139, USA}

\author[0000-0002-9238-9521]{David Sanmartim}
\affiliation{Carnegie Observatories, Las Campanas Observatory, Casilla 601, La Serena, Chile} 

\author[0000-0003-2445-3891]{Matthew R. Siebert}
\affiliation{Space Telescope Science Institute, 3700 San Martin Drive, Baltimore, MD 21218, USA}

\author[0000-0003-1772-0023]{Thaisa Storchi-Bergmann}
\affiliation{Departamento de Astronomia - IF, Universidade Federal do Rio Grande do Sul, CP 150501, 91501-970 Porto Alegre, RS, Brazil}

\author[0000-0001-9191-9837]{Marianne Vestergaard}
\affiliation{Steward Observatory, University of Arizona, 933 North Cherry Avenue, Tucson, AZ 85721, USA}
\affiliation{DARK, The Niels Bohr Institute, University of Copenhagen, Universitetsparken 5, DK-2100 Copenhagen, Denmark}

\author[0000-0001-9449-9268]{Jian-Min Wang}
\affiliation{Key Laboratory for Particle Astrophysics, Institute of High Energy Physics, Chinese Academy of Sciences, 19B Yuquan Road,\\ Beijing 100049, People's Republic of China}
\affiliation{School of Astronomy and Space Sciences, University of Chinese Academy of Sciences, 19A Yuquan Road, Beijing 100049, People's Republic of China}
\affiliation{National Astronomical Observatories of China, 20A Datun Road, Beijing 100020, People's Republic of China}

\author[0000-0002-5205-9472]{Tim Waters}
\affiliation{Theoretical Division, Los Alamos National Laboratory, Los Alamos, NM, USA}

\author[0000-0003-0931-0868 ]{Fatima Zaidouni}
\affiliation{MIT Kavli Institute for Astrophysics and Space Research, Massachusetts Institute of Technology, Cambridge, MA 02139, USA}

\begin{abstract}
The AGN STORM 2 campaign is a large, multiwavelength reverberation mapping project designed to trace out the structure of \mrk\ from the inner accretion disk to the broad emission line region and out to the dusty torus.  As part of this campaign, \swift\ performed daily monitoring of \mrk\ for approximately 15 months, obtaining observations in X-rays and six UV/optical filters.  The X-ray monitoring shows that \mrk\ was in a significantly fainter state than in previous observations, with only a brief flare where it reached prior flux levels.  The X-ray spectrum is heavily obscured.  The UV/optical light curves show significant variability throughout the campaign and are well correlated with one another, but uncorrelated with the X-rays.  Combining the Swift UV/optical light curves with Hubble UV continuum light curves, we measure interband continuum lags, $\tau(\lambda)$, that increase with increasing wavelength roughly following $\tau(\lambda) \propto \lambda^{4/3}$, the dependence expected for a geometrically thin, optically thick, centrally illuminated disk.  Modeling of the light curves reveals a period at the beginning of the campaign where the response of the continuum is suppressed compared to later in the light curve -- the light curves are not simple shifted and scaled versions of each other.  The interval of suppressed response corresponds to a period of high UV line and X-ray absorption, and reduced emission line variability amplitudes. We suggest that this indicates a significant contribution to the continuum from the broad line region gas that sees an absorbed ionizing continuum.
\end{abstract}

\section{Introduction}

With just a handful of exceptions \citep{EHT_M87, gravity_3c273, gravity_iras09149, gravity_3783}, the angular size of the inner regions of active galactic nuclei (AGNs) are too small to be spatially resolved.  Piecing together the geometry and kinematics of the inner regions of AGNs therefore requires the use of additional techniques.  Variability studies are particularly powerful, with small size scales achievable through high time resolution. In particular, reverberation mapping \citep{blandmckee82, peterson14} uses time lags, $\tau$, between  light curves  at different wavelengths to determine the size scale, $R \approx c\tau$, of emitting regions.  By observing continuum and emission lines from X-rays through the near-infrared (IR) one can probe the locations of the X-ray corona, accretion disk, broad-line region (BLR), dusty torus and beyond \citep[see][for a recent review]{cackett21}.  

A big step forward in reverberation mapping studies was the large, coordinated, multiwavelength AGN STORM campaign on NGC~5548 \citep{derosa15}.  During this campaign the Hubble Space Telescope (\hst) performed daily ultraviolet (UV) spectroscopic monitoring of NGC~5548, with contemporaneous monitoring by \swift\ \citep{edelson15} plus ground-based photometry \citep{fausnaugh16} and spectroscopy \citep{pei17}.  AGN STORM revealed a number of interesting and surprising results, including identification of a period where all the emission lines and high-ionization absorption lines decoupled from the continuum variations \citep[dubbed the ``BLR holiday'';][]{goad16}, suggesting the presence of a variable, obscuring disk wind \citep{dehghanian19a,dehghanian19b}.  

AGN STORM also showed clear wavelength-dependent continuum reverberation lags  spanning approximately 1000 -- 10000~\AA\ and roughly following $\tau \propto \lambda^{4/3}$ \citep{edelson15,fausnaugh16}, the wavelength-dependence expected for a standard \citep[e.g.,][]{shakurasunyaev} geometrically thin, optically thick accretion disk \citep[e.g.,][]{collier99,cackett07}. However, the magnitude of the lags was a factor of $\sim$3 larger than predicted using analytical models and reasonable estimates of the mass accretion rate  \citep{edelson15,fausnaugh16}, a problem that is now commonly seen in continuum reverberation mapping experiments \citep[e.g.,][]{jiang17,cackett18, cackett20, fausnaugh18, mudd18, edelson19,pozonunez19,jha22,guo22}.  A similar disk size problem was first identified from microlensing studies of gravitationally lensed quasars \citep[e.g.,][]{morgan10}.  \citet{kammoun19,kammoun21a,kammoun21b} address the (apparent) size discrepancy using more detailed models of a centrally illuminated, reverberating thin disk, though in the absence of any BLR contribution to the continuum emission, which photoionization models predict must be present at some level.

A common model for continuum reverberation lags is the lamppost model, where variations in X-ray emission drive the observed variability at longer wavelengths through thermal reprocessing in the disk. However, the X-ray variations in NGC~5548 were seen to not be simply related to the variability at longer wavelengths \citep{starkey17, gardnerdone17}.  Similarly, observations of other objects also show that the X-ray to UV correlation is typically weaker than the UV to optical correlation \citep[e.g.,][]{edelson19,cackett20,hernandezsantisteban20}.  Dynamic variability in the X-ray source can potentially explain the moderate X-ray to UV correlation \citep{panagiotou22}.  However, in some cases the X-ray and UV are not correlated at all \citep[e.g.,][]{schimoia15,buisson18,morales19}, presenting a challenge to the simplest reprocessing scenario.

Another complication seen in AGN STORM was that the $U$ band showed a lag in excess of an extrapolation of the other UV/optical bands \citep{edelson15,fausnaugh16}.  This can be interpreted as the BLR producing significant bound-free and free-free continuum emission, the spectrum of which should peak at the Balmer jump, which lies in the $U$ band \citep{koristagoad01,koristagoad19, lawther18, netzer20, netzer22}. $U$-band excess lags are almost universally seen \citep{mchardy14,mchardy18,edelson17,edelson19,cackett20,hernandezsantisteban20}, and spectral-timing analysis revealed a broad lag excess and discontinuity at the Balmer jump in NGC 4593 \citep{cackett18}.  While the BLR diffuse continuum peaks locally at the Balmer and Paschen jumps, it contributes broadly across the UV/optical \citep[e.g.,][]{koristagoad01,netzer22}.  More advanced time lag analysis also suggests a significant contribution from a more distant reprocessor throughout the UV and optical as well \citep{chelouche19,cackett22}. A contribution to the lags from the BLR diffuse continuum emission could account for a significant portion (possibly all) of the apparent disk size disparity.

The AGN STORM 2 project is the next large, coordinated, multiwavelength reverberation mapping campaign.  It is built around a large \hst\ program \citep{peterson20} to monitor the nearby Seyfert 1 galaxy \mrk\ ($z = 0.031455$) every other day for approximately 15 months.  Additional high-cadence contemporaneous monitoring data were obtained by \swift, NICER, optical and near-IR ground-based photometry and spectroscopy plus a smaller number of deeper observations by XMM-Newton and NuSTAR. \mrk\ has a black hole mass of $M = 3.85\times10^7$~M$_\odot$ \citep{bentzkatz15}, and is accreting at an Eddington ratio of approximately $\dot{m}_{\rm E} = 0.2$.  \citet{kara21}, hereafter Paper I, describes the campaign in detail and presents the results from the first three months of the project.  Initial observations at the beginning of the campaign discovered unexpected and variable obscuration in both the UV and X-ray (\citealt{kara21}, and also \citealt{miller21} for an independent analysis of the X-ray obscuration).  Despite this, both UV/optical continuum reverberation and emission line reverberation were still observed.  \citet[hereafter Paper II]{homayouni23} presents the \hst\ observations and UV emission line reverberation from the full campaign.  Paper III \citep{partington23} presents a detailed analysis of the X-ray spectral variability using NICER.  Here, in Paper IV, we present the \swift\ observations from the full campaign, and study the UV/optical continuum variability and reverberation.  

In Section~\ref{sec:data} we describe the data reduction, and in Section~\ref{sec:analysis} we present our analysis, including the X-ray variability, UV/optical continuum reverberation, and an analysis of the variable UV/optical spectral energy distribution (SED).  Finally, we discuss the implications of our results in Section~\ref{sec:discuss}.

\section{Data Reduction}\label{sec:data}

The Neil Gehrels Swift Observatory (hereafter \swift) monitored \mrk\ daily, concurrently with \hst , with 1 ksec observations for $\sim$15 months from 2020 November 22 to 2022 February 24.  Data were obtained both through a target of opportunity request and a \swift\ Key Project (proposal number 1720084, PI: Cackett).  \swift\ monitoring is continuing as part of an extended campaign (proposal 1821087, PI: Gelbord), but we limit this analysis to those data that overlap with the \hst\ project. There are occasional gaps of a few days in the  \swift\ monitoring caused by limited visibility during orbital pole observing constraints. Toward the very end of the campaign \swift\ entered a safe mode due to the loss of a reaction wheel. This resulted in no observations being performed for 1 month between 2022 January 18 and 2022 February 18.

\begin{figure*}
\centering
\includegraphics[trim={0 0.65cm 0 0},clip,width=0.9\textwidth]{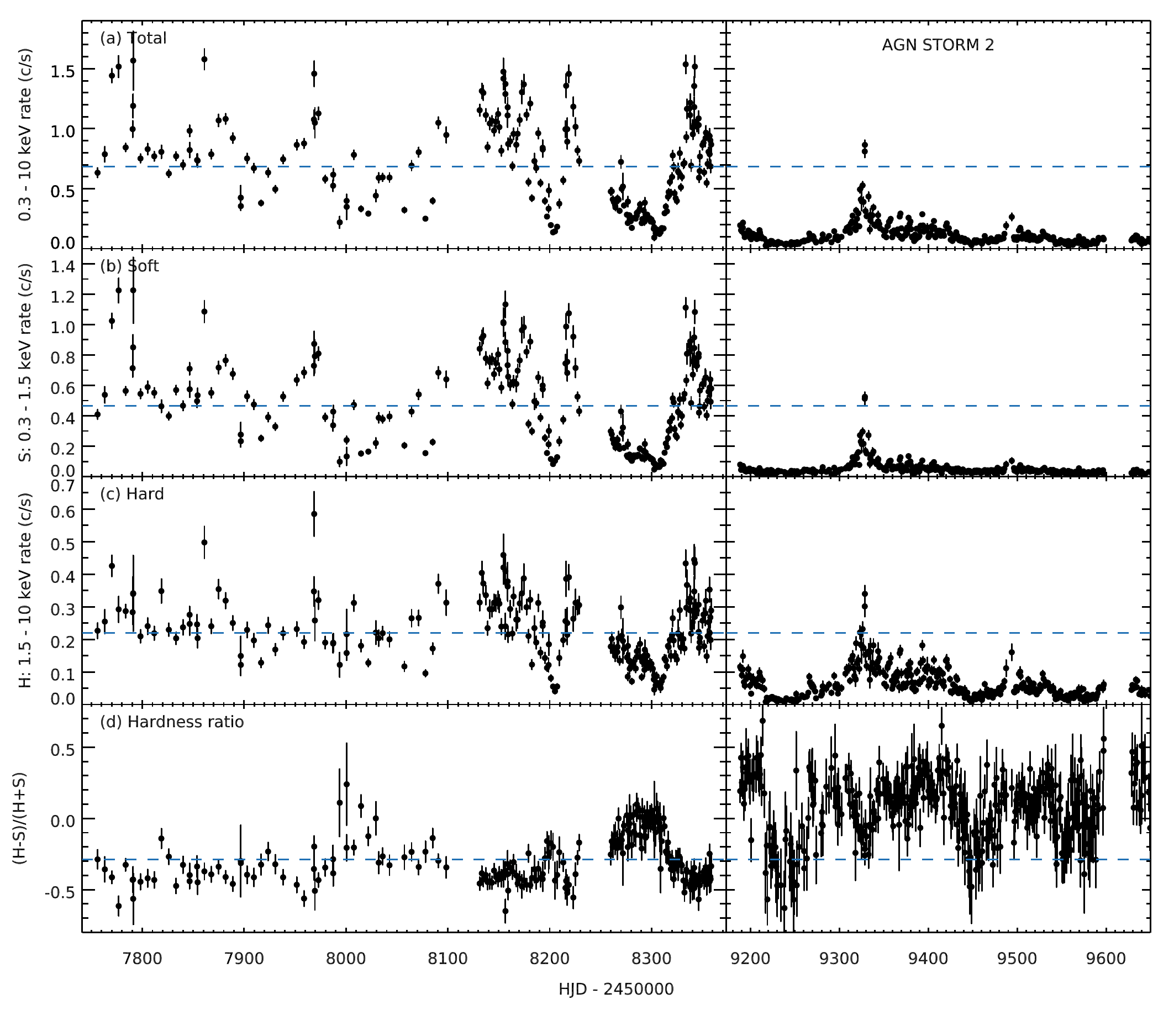}
\caption{\swift\ X-ray light curve of \mrk\ in the (a; top) total: 0.3 -- 10~keV, (b; upper middle) soft: 0.3 -- 1.5~keV, and (c; lower middle) hard: 1.5 -- 10~keV bands. Panel (d; bottom) shows the hardness ratio.  Left-hand panels show the data from \citet{morales19}, while the right-hand panels show the AGN STORM 2 campaign.  There is a gap of $\sim$830 days between the two campaigns.  Blue dashed lines indicate the historical average fluxes calculated from all \swift\ data taken before this campaign.  A dramatic drop in flux is seen for all bands during AGN STORM 2. }
\label{fig:xraylc}
\end{figure*}

The X-ray light curves are produced using the \swift/XRT data products generator\footnote{\url{https://www.swift.ac.uk/user_objects/}} \citep{evans07,evans09}, with background-subtracted count rates obtained in the 0.3 -- 1.5~keV (soft, $S$) and 1.5 -- 10~keV (hard, $H$) bands per snapshot (spacecraft orbit).  To crudely trace spectral changes, we calculate a hardness ratio, $HR = (H-S)/(H+S)$.  Fig.~\ref{fig:xraylc} shows the X-ray light curves and hardness ratio during the AGN STORM 2 campaign (right panels) and during the previous $\sim$15 months of monitoring by \citet{morales19} for comparison. A dramatic drop in the X-ray count rates, a factor $\sim$ 6 on average, is seen during the AGN STORM 2 campaign compared to the historical average.  The X-ray count rates are given in Table~\ref{tab:lc}.  Dates are given as HJD $-$ 2,450,000 throughout the paper.

\begin{figure*}
\centering
\includegraphics[trim={0 1.5cm 2cm 0},clip, width=0.9\textwidth]{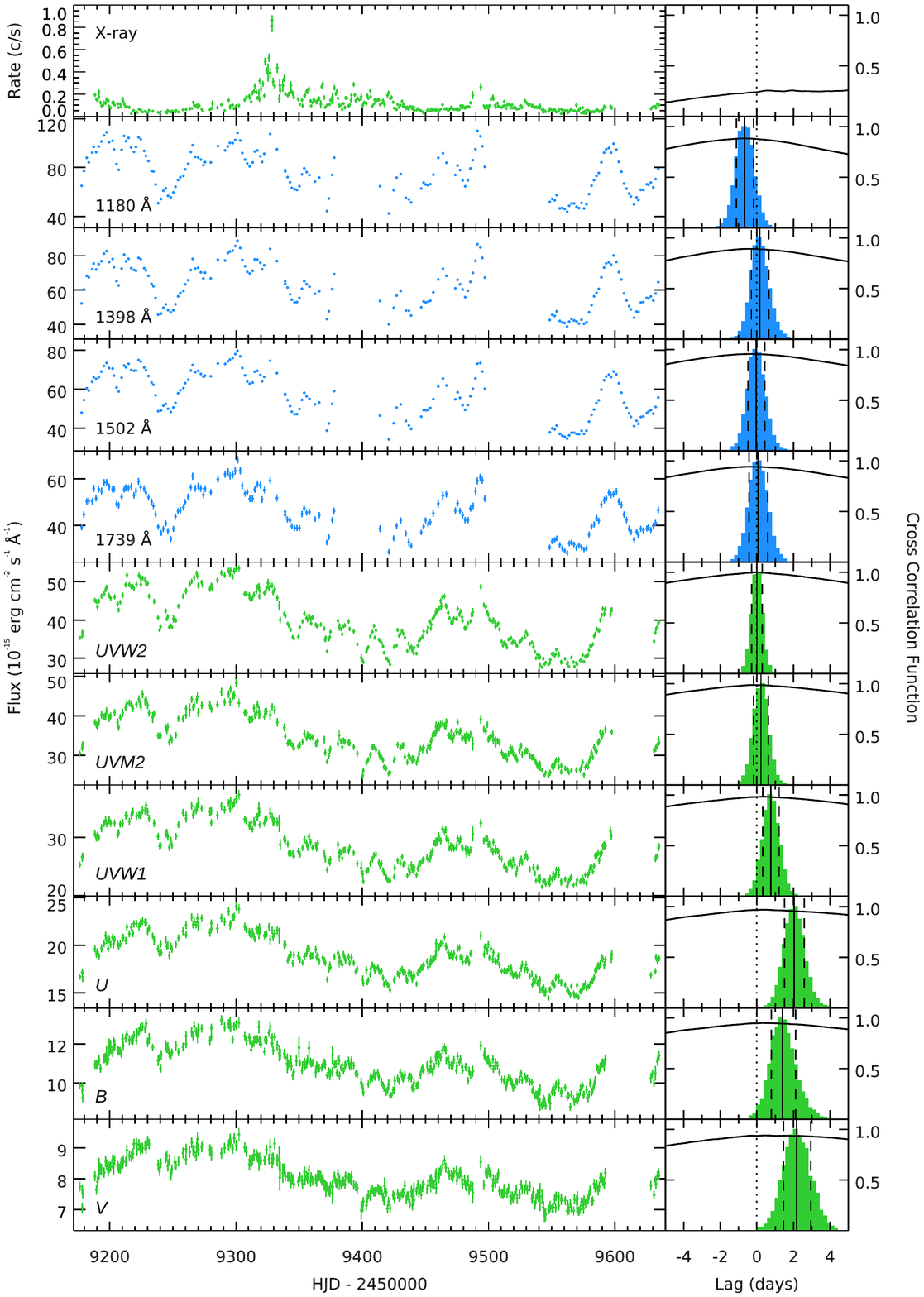}
\caption{{\it Left:} \swift\ (green) and \hst\ (blue) light curves of \mrk\ during the AGN STORM 2 campaign. {\it Right:} Cross-correlation functions calculated with respect to the light curve in the \swift/{\it UVW2} band (solid lines). Normalized histograms show the ICCF centroid distribution, with the vertical solid and dashed lines showing the lag centroid and 1$\sigma$ uncertainty range, respectively.}
\label{fig:alllc}
\end{figure*}

The UVOT analysis follows the procedure outlined in previous work  \citep{edelson15,edelson17,edelson19, cackett20, hernandezsantisteban20}.  The data were processed with \verb\HEASOFT\ v6.29.  The flux of \mrk\ is measured for each epoch and filter using the \verb\UVOTSOURCE\ tool.  We use a circular source region with a 5\arcsec\ radius, and background annulus from 40--90\arcsec.  Any stars that fall within the background annulus are excluded.  Data quality tests are used to screen out observations bearing evidence of target tracking errors or extended point-spread functions (PSFs), eliminating between 5 and 21 measurements per filter.  We also use detector masks to screen out measurements that are likely to be significantly affected by areas of the detector with reduced sensitivity.  The process of determining the detector mask is described in \citet{hernandezsantisteban20}.  In this work we found that a less aggressive detector mask could be used as compared to previous work.  Those prior efforts used the equivalent of the ``Mid'' small scale sensitivity (SSS) maps whereas we opt for the ``Low'' SSS masks that are now available as part of the \swift\ CALDB \citep{breeveld22}.  This masking process rejects between 3.4\% ({\it B}) to 18.4\% ({\it UVM2}) of the images. The resulting light curves contain between 346 ({\it V}) and 292 ({\it UVM2}) epochs.  The \swift\ UVOT light curves are given in Table~\ref{tab:lc} and are shown in Fig.~\ref{fig:alllc}.

In addition to the Swift X-ray and UV/optical light curves, we also include the UV continuum light curves from HST in our analysis.  The HST data analysis is described in Paper II \citep{homayouni23} and the data can be obtained from the Mikulski Archive for Space Telescopes (MAST) at the Space Telescope Science Institute via \dataset[doi:10.17909/n734-k698]{https://doi.org/10.17909/n734-k698}.

\begin{deluxetable}{cccc}
\label{tab:lc}
\tablecaption{Swift photometry of \mrk}
\tablewidth{0pt}
\tablehead{
 \colhead{HJD $-$ 2,450,000} & \colhead{Filter/Band} & \colhead{Rate/Flux} & \colhead{Uncertainty} }
\decimalcolnumbers
\startdata
9188.412   &   S &  0.079  &      0.010  \\
9189.476  &  S  &  0.044   & 0.008       \\
9190.406   &    S & 0.054  & 0.008       \\
9191.402   &  S  & 0.067   & 0.010      \\
9191.600   & S  &  0.070    & 0.010      \\
\dots & \dots & \dots & \dots
\enddata
\tablecomments{This table is published online in its entirety in the machine-readable format.  A portion is shown here for guidance regarding its form and content. S is the X-ray rate in the 0.3 -- 1.5 keV band, while H is the X-ray rate in the 1.5 -- 10 keV band.  The X-ray rates are given as count rates, while the Swift/UVOT fluxes have units of $10^{-15}$~erg~cm$^{-2}$~s$^{-1}$ \AA$^{-1}$.}
\end{deluxetable}

\section{Analysis and Results} \label{sec:analysis}
\subsection{X-ray variability}

The X-ray light curves (Fig.~\ref{fig:xraylc}) show a significant change in both count rate and hardness compared to the previous \swift\ monitoring of \mrk.  The count rates during AGN STORM 2 are on average a factor of 6 lower.  Variability, especially in the soft band, is generally low aside from a notable and dramatic flare  peaking on day 9329.  During the flare, the 0.3 -- 10 keV count rate increases by a factor of $\sim$10 compared to the mean rate around day 9300, briefly (for a few days) exceeding the historical averages in both the soft and hard bands.  Increases in count rate of this magnitude over this timescale are seen in earlier monitoring of \mrk\ \citep{morales19}, but it is more pronounced here given the overall lower fluxes and variability throughout the rest of the campaign and the higher monitoring cadence.  Previous faint X-ray states for \mrk\ have been observed -- longer term X-ray monitoring presented in \citet{winter11} shows that ROSAT observations from the early 1990s found \mrk\ a factor of 40 fainter than a 2009 XMM-Newton observation.  For comparison, the 0.4 -- 8 keV X-ray flux observed by NICER ranged from $9\times10^{-13}$ to $2\times10^{-11}$~erg~cm$^{-2}$~s$^{-1}$ during STORM 2 \citep{partington23}, while ROSAT observed a minimum of $8\times10^{-13}$~erg~cm$^{-2}$~s$^{-1}$ (0.1 -- 2.4 keV) and XMM-Newton a maximum of $3.3\times10^{-11}$~erg~cm$^{-2}$~s$^{-1}$ \citep[0.3 -- 10 keV;][]{winter11}.

\begin{figure}
\centering
\includegraphics[width=0.9\columnwidth]{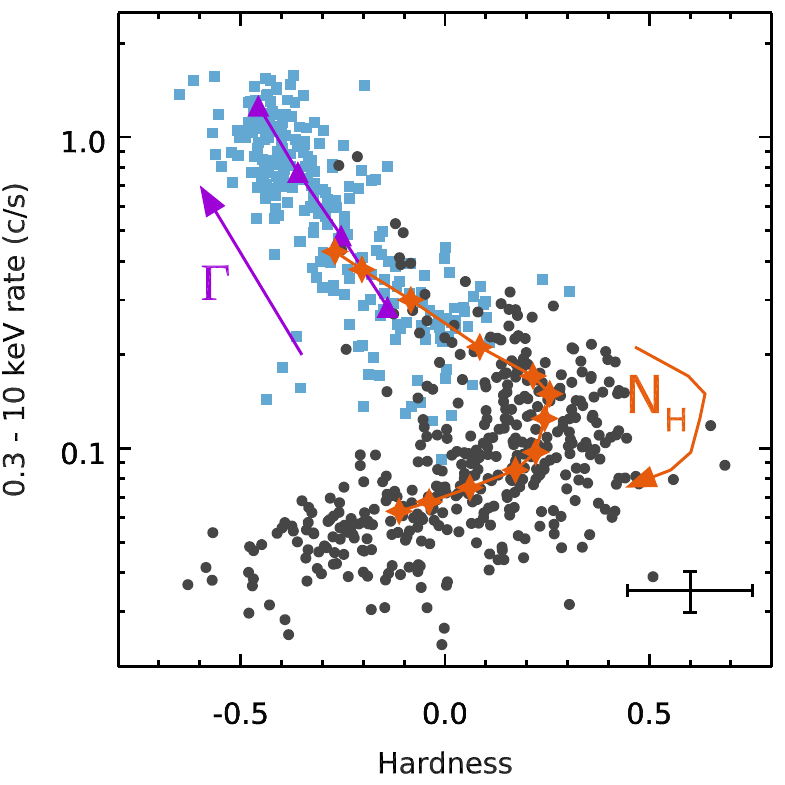}
\caption{X-ray hardness-intensity diagram for Mrk 817.  Black circles show data taken during the AGN STORM 2 campaign, while blue squares show data taken prior to this campaign.  The variability patterns are clearly different.  Error bars are omitted for clarity, but a representative error bar is shown.  Purple triangles show the evolution of a power-law that is softer-when-brighter.  Orange stars shown the evolution of a partially covered power-law with increasing column density from $N_{\rm H} = 2.5\times10^{21}$~cm$^{-2}$ to $2.5\times10^{23}$~cm$^{-2}$.  Arrows indicate the direction in which the parameters increase.}
\label{fig:hid}
\end{figure}

The average background-subtracted 0.3 -- 10 keV count rate during AGN STORM 2 is around 0.1 c~s$^{-1}$, leading to typically less than 100 counts in each spectrum.  With so few counts, we do not perform individual spectral fits.  Instead, we examine the X-ray spectral variability using an X-ray hardness-intensity diagram (HID), comparing the hardness ratio to the 0.3 -- 10~keV count rate (see Fig.~\ref{fig:hid}).  Data from AGN STORM 2 mostly fall in a different part of the HID than earlier observations.  
 
Previous X-ray variability studies of Mrk~817 have shown a softer-when-brighter behavior \citep{winter11,morales19}.  To demonstrate how this changes the hardness ratio, we used {\sc xspec} \citep{arnaud96} to calculate the expected count rates and hardness for power-law spectra with power-law indices  ranging from $\Gamma = 2.2$ to 1.6 (in steps of 0.2), assuming Galactic absorption of $N_{\rm H} = 1.5\times10^{20}$~cm$^{-2}$ \citep{dickey90} and varying the power-law normalization to roughly match the observed trend on the HID (purple triangles in Fig.~\ref{fig:hid}). Note that a change in normalization simply shifts the points up and down, while a change in power-law index alters the hardness ratio.  The softer-when-brighter trend matches the historical \swift\ data, as expected given the spectral fitting performed by \citet{morales19}.  But, a simple power-law with just Galactic absorption does not match the rest of the variability seen in the HID.

Since fits to higher quality XMM-Newton, NuSTAR and NICER spectra \citep{kara21,miller21,partington23} show the spectrum is highly obscured, we also investigate the impact of an obscurer on the hardness.  Here, we assume $\Gamma = 2.0$, a power-law normalization of $5\times10^{-3}$~photons~keV$^{-1}$~cm$^{-2}$~s$^{-1}$ at 1 keV, and the same Galactic absorption.  However, we also include a partial covering absorber (zxipcf in {\sc xspec}), with a covering factor of 0.9 and an ionization parameter of $\log\xi = 1.0$.  We vary the column density of the absorber from $N_{\rm H} = 2.5\times10^{21}$~cm$^{-2}$ to $2.5\times10^{23}$~cm$^{-2}$.  The results are shown as orange stars in Fig.~\ref{fig:hid}.  At first, the increasing $N_{\rm H}$ leads to a decreasing count rate and a hardening of the spectrum.  But, above approximately $N_{\rm H} = 3.5\times10^{22}$~cm$^{-2}$ the spectrum softens because the increasing column density begins to decrease the count rates in the hard band above 1.5 keV.  Thus, the softest spectra at the lowest observed count rates are due to the highest absorbing column.  

While these models are just for illustration, it indicates that the main trend in the HID during AGN STORM 2 is consistent with significant changes in the obscurer, but to explain the full HID requires variability in the intrinsic flux as well.  This is explored in significantly more detail in Paper III \citep{partington23}, which tracks the change in the X-ray obscurer using NICER spectral monitoring.  The hardness ratio changes are similar to what is seen in NGC~5548 where changes in both intrinsic flux and obscuration are also seen \citep{mehdipour16}.

\subsection{Time Series Analysis}

All the light curves show clear variability throughout the campaign (see Fig.~\ref{fig:alllc}), but, the X-ray variability is quite different from the UV/optical.  The UV/optical light curves show strong peaks and troughs that are seen at all wavelengths, with the longer wavelength variability features smoother than the shorter wavelength ones.  This UV/optical behavior is typical of other continuum reverberation mapping campaigns.  The X-ray behavior, however, is quite atypical.  Usually there is some level of correlation between the X-ray and UV/optical \citep[e.g.,][]{mchardy18}, and the X-rays vary more rapidly than the UV/optical \citep[e.g.,][]{gardnerdone17,starkey17}.  We observe a difference in \mrk\ -- there is no significant correlation and much of the structure in the UV/optical variability is not seen in the X-rays.

Before performing a time lag analysis, we qualitatively compare the X-ray, {\it UVW2} and HST 1180 \AA\ light curves.  The first illustrates the lack of correlation with the X-ray and the second illustrates that the {\it UVW2} light curve is not simply a shifted and smoothed version of the 1180 \AA\ light curve.  We then measure the time lags between the different UV/optical bands using a variety of analysis techniques.  We first use the standard interpolation cross-correlation function (ICCF) method \citep{gaskellpeterson87,white94,petersonetal04}, then we apply more sophisticated approaches using the {\sc Javelin} \citep{zu11,zu13} and {\sc PyROA} \citep{donnan21} techniques. We include the four \hst\ continuum light curves from Paper II in addition to the six \swift\ UVOT light curves measured here.  The central wavelengths for all the bands are given in Table~\ref{tab:lags}.  The \swift\ light curves have a higher cadence than the \hst\ light curves, so we use the \swift/{\it UVW2} band as the reference.

\subsubsection{X-ray vs {\it UVW2}}

\begin{figure*}
\centering
\includegraphics[width=0.95\textwidth]{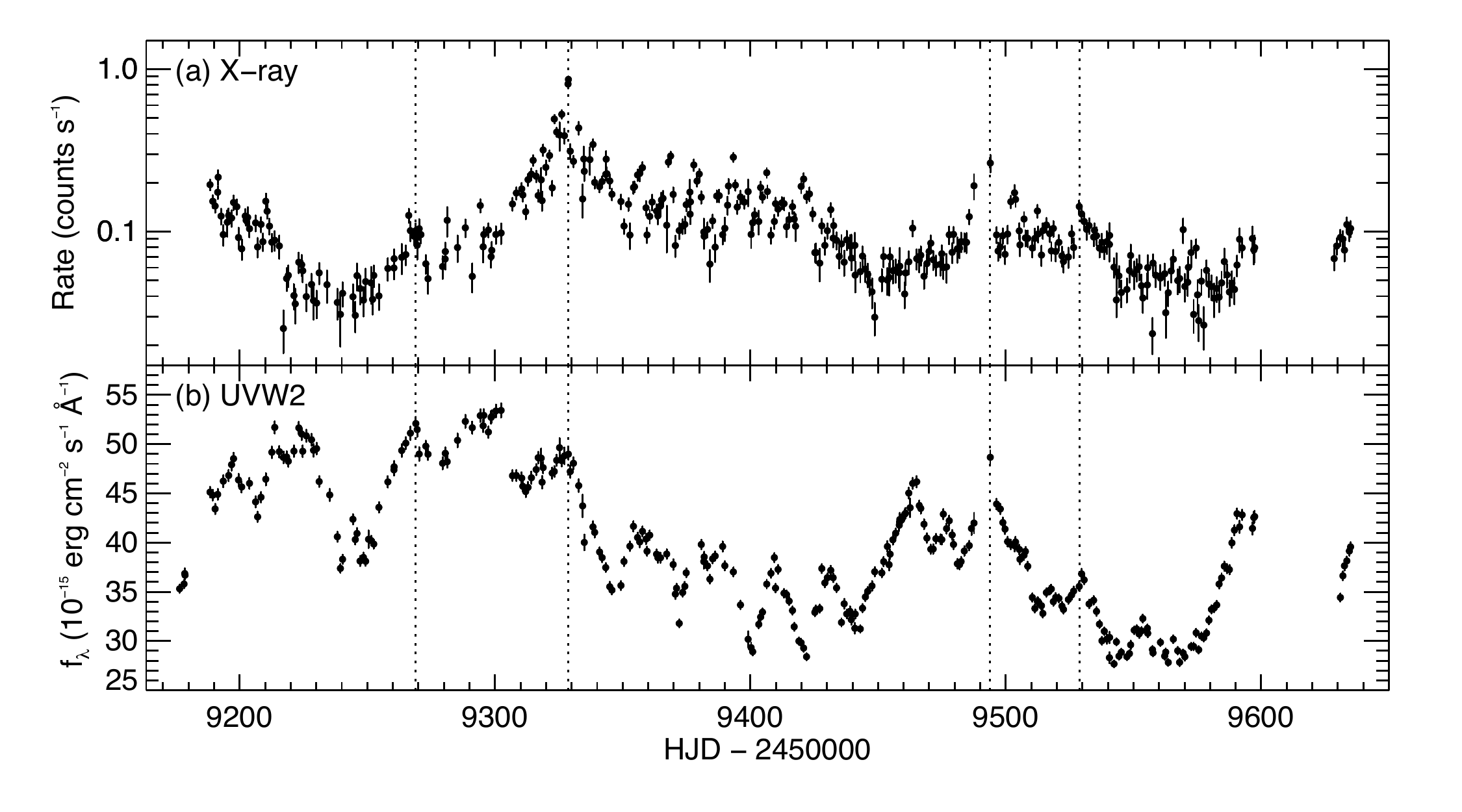}
\caption{A detailed comparison of the (a) X-ray and (b) {\it UVW2} light curves of \mrk. The X-ray light curve is on a log scale to better highlight the variability at low count rates.  Generally, the features of the two light curves do not match up, with the exception of a few peaks indicated by vertical dotted lines.}
\label{fig:xrayw2}
\end{figure*}

In the top panel of Fig.~\ref{fig:alllc}, it is immediately apparent that there is little correlation between the X-ray and UVOT light curves. This lack of correlation was seen previously in \mrk\ by \citet{morales19} and in Paper I.  We calculate the cross-correlation function (CCF) between the X-ray and {\it UVW2} light curves. The maximum correlation coefficient for a lag between $-20$ and 20~days is just 0.33 and no lag can be determined between the X-ray and {\it UVW2} bands.  To better display the relationship between the X-ray and {\it UVW2} light curves we show a larger version of just these two light curves in Fig.~\ref{fig:xrayw2}, with the X-ray count rates on a log scale to better highlight the variability at low count rates.  While there are some peaks in the X-ray light curve that correspond to a peak in the {\it UVW2} light curve, as indicated by vertical dotted lines, the {\it UVW2} variability does not match the variability in the X-rays generally. There appears to be better agreement between the {\it UVW2} and X-ray light curves after around day 9310.  Performing a CCF on just that portion of the light curves does give a better maximum correlation coefficient of $\sim$0.5, but there is no clear peak and the correlation coefficient is still too low to determine a lag.

\subsubsection{1180~\AA\ vs {\it UVW2}}

\begin{figure*}
\centering
\includegraphics[width=0.95\textwidth]{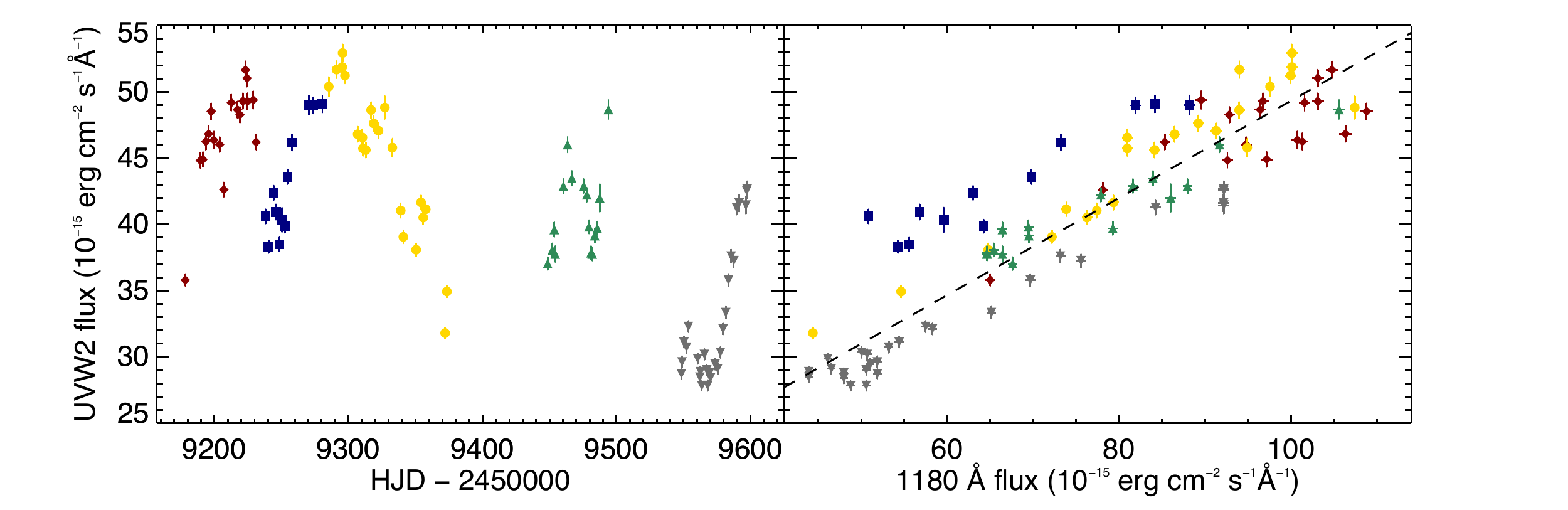}
\caption{{\it Left:} {\it UVW2} light curve divided into the five time segments identified in Homayouni et al., submitted (only points within the time segment and within $\pm0.5$ days of a 1180~\AA\ data point are shown). {\it Right:} The correlations between the {\it UVW2} and 1180~\AA\ fluxes for the different time segments. Colors and symbols match the time segments on the left. The dashed line is the best-fitting linear relation to the whole dataset.}
\label{fig:1180W2}
\end{figure*}

\begin{figure*}
\centering
\includegraphics[width=0.95\textwidth]{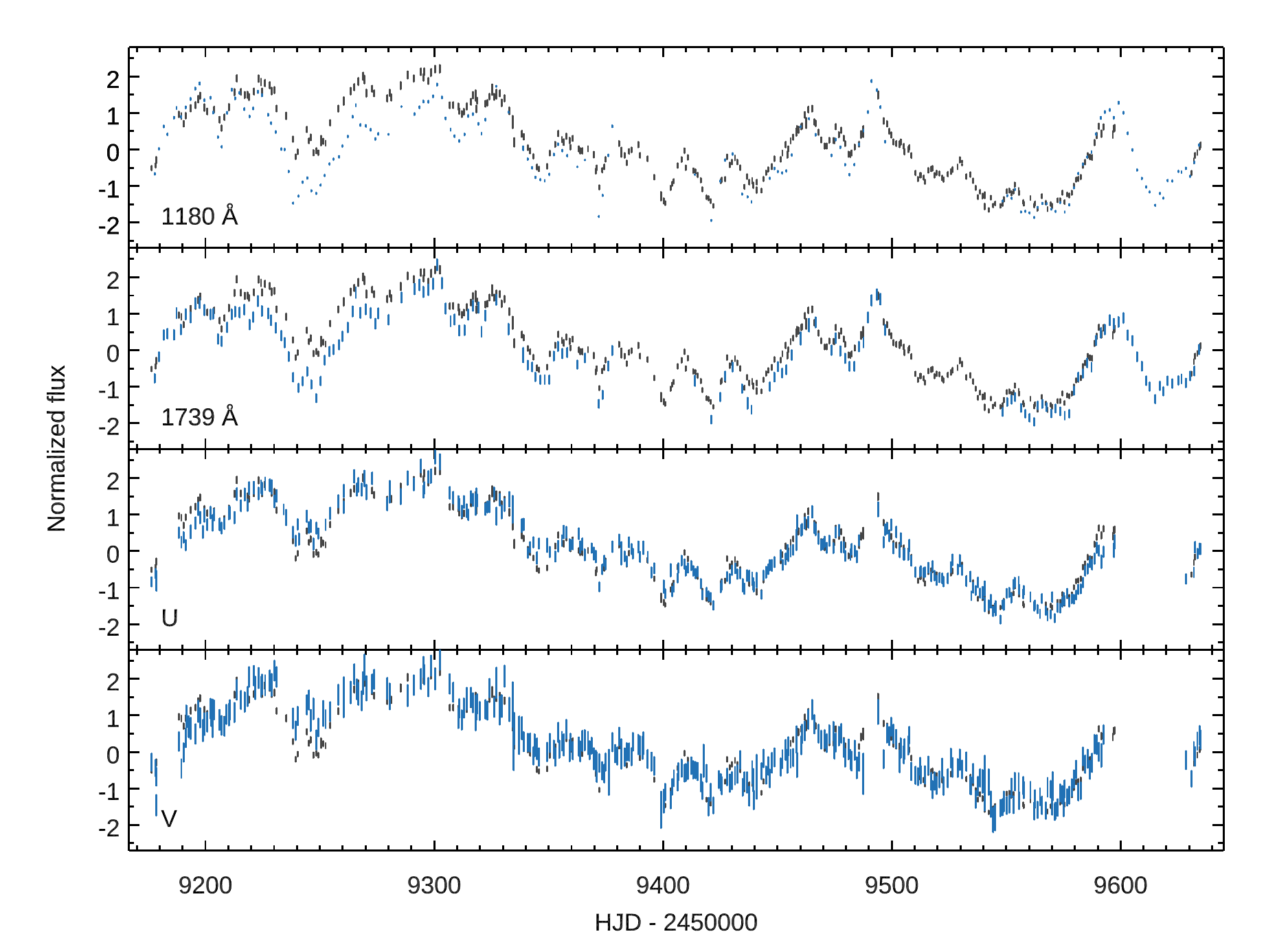}
\caption{A comparison of the normalized {\it UVW2} light curve (black) with normalized light curves (blue) at 1180 \AA\ (top), 1739 \AA, U and V (bottom).  All light curves are normalized by subtracting the median and dividing by the standard deviation of the full light curve.  No time shift was applied.}
\label{fig:W2comp}
\end{figure*}

An initial visual inspection of the light curves in Fig.~\ref{fig:alllc} shows a different long-term trend in the two shortest wavelength \hst\ light curves (1180~\AA\ and the 1398~\AA) compared to the longer wavelength Swift UVOT light curves. This is most evident when comparing the light curve peaks near days 9300 and 9490.  In the shortest wavelength light curves these two peaks reach approximately the same flux (the second is brightest in 1180~\AA), while at longer wavelengths the second peak is substantially fainter. Homayouni et al. (submitted) found that the UV emission lines could not simply be fit by shifting, smoothing, and scaling the 1180~\AA\ continuum during AGN STORM 2.  This is not unexpected at least from a photoionization point of view \citep{goad93,koristagoad04,goadkorista15}, since local gas physics suggest few lines (with the possible exception of \ion{He}{2} and Fe~K$\alpha$) will respond linearly to variations in the incident ionizing continuum flux, and even a `linearized' echo model will break down in the face of large amplitude continuum variations (peak to peak variations of factors of several or more). However, splitting the light curves into different segments they found that the emission-line lags differed in each segment.

To assess whether the different long-term continuum trends we observe here may be associated with the time segments identified in Homayouni et al. (in prep), in Fig.~\ref{fig:1180W2} we compare the 1180~\AA\ flux to the {\it UVW2} flux, for times when the {\it UVW2} observation was obtained within $\pm0.5$ days of the 1180~\AA\ observation.  The different time segments show approximately the same slope (indicating a similar response to changes in ionizing flux), but there appears to be a vertical offset between them -- this is most apparent between the second (blue) and last (gray) time segments.  Aside from the first $\sim$150 days, all segments behave in a similar way.

We investigate this further by normalizing each of the light curves for direct comparison to the {\it UVW2} light curve by subtracting the median and dividing by the standard deviation. In this format, light curves that are simply scaled versions of each other should overlap, and any deviations from this become apparent.  Fig.~\ref{fig:W2comp} compares several normalized light curves to the normalized {\it UVW2} light curve.  For the shortest wavelength bands we see a discrepancy between the light curves during the first $\sim$150 days.  At later times, the light curves are generally consistent with being scaled versions of each other.  The variations are well correlated over the first part of the light curve, while they cannot be described by the same scaling.  This is most prominent comparing the shortest wavelength (1180~\AA) light curve to {\it UVW2},  but even between the \swift\ bands a discrepancy can be seen at the double trough around days 9240 -- 9250.  The troughs are significantly deeper at the shortest wavelengths.

As we discuss later, these different long-term trends present a problem for modeling the light curves, and therefore we perform lag analysis on both the original and detrended light curves.  Detrending is the process of subtracting a long-timescale trend from the light curve \citep[e.g.,][]{welsh99}.  We do this by subtracting a running Gaussian average with a width of $\sigma = 20$ days for each light curve independently.   We initially tried a linear detrending, but this did not fully solve the scaling mismatch early in the light curve.  We experimented with a range of different widths for the Gaussian, and chose the broadest width Gaussian that allows the 1180~\AA\ and {\it UVW2} light curves to be well-matched.

\subsubsection{ICCF}

We first calculate the interband lags using the ICCF technique. Briefly, we calculate  the CCF between each of the light curves and the {\it UVW2} reference light curve.  The CCF is calculated from $-20$ to 20 days (but only shown from $-4.5$ to 4.5 days in Fig.~\ref{fig:alllc} for clarity).  The flux randomization/random-subset sampling (FR/RSS) method \citep[as implemented by][]{petersonetal04} is used to determine the uncertainties in the lags, with 10,000 light curve realizations.  This both randomizes the flux at each epoch (FR) assuming a Gaussian distribution with a mean and standard deviation equal to the observed flux and its measurement uncertainty, and samples a subset of on average $\sim$2/3 of the data with replacement (RSS). For each realization the CCF is calculated and the peak and centroid of the CCF are determined, with the centroid calculated using only points that are greater than 0.8 of the maximum CCF value, $R_{\rm max}$.  The median and 68\% confidence intervals of the centroid and peak distributions are used to determine the lag peak ($\tau_{\rm peak}$), centroid ($\tau_{\rm cent}$) and their associated uncertainties.

The \swift\ light curves have a higher cadence than the \hst\ light curves and so lead to better constrained lags.  Of the \swift\ UVOT bands, the {\it UVW2} band has the highest variability amplitude and shortest wavelength, and so we adopt it as the reference band relative to which we measure the lags of the other bands.   The lags, along with the variability amplitude \citep[$F_{\rm var}$;][]{vaughan03}, the maximum correlation coefficient ($R_{\rm max}$), and the filter central wavelengths are given in Table~\ref{tab:lags}.  All lags here are quoted in the rest frame, assuming a redshift of $z = 0.031455$ \citep{strauss88}.

\begin{figure*}
\centering
\includegraphics[width=14cm]{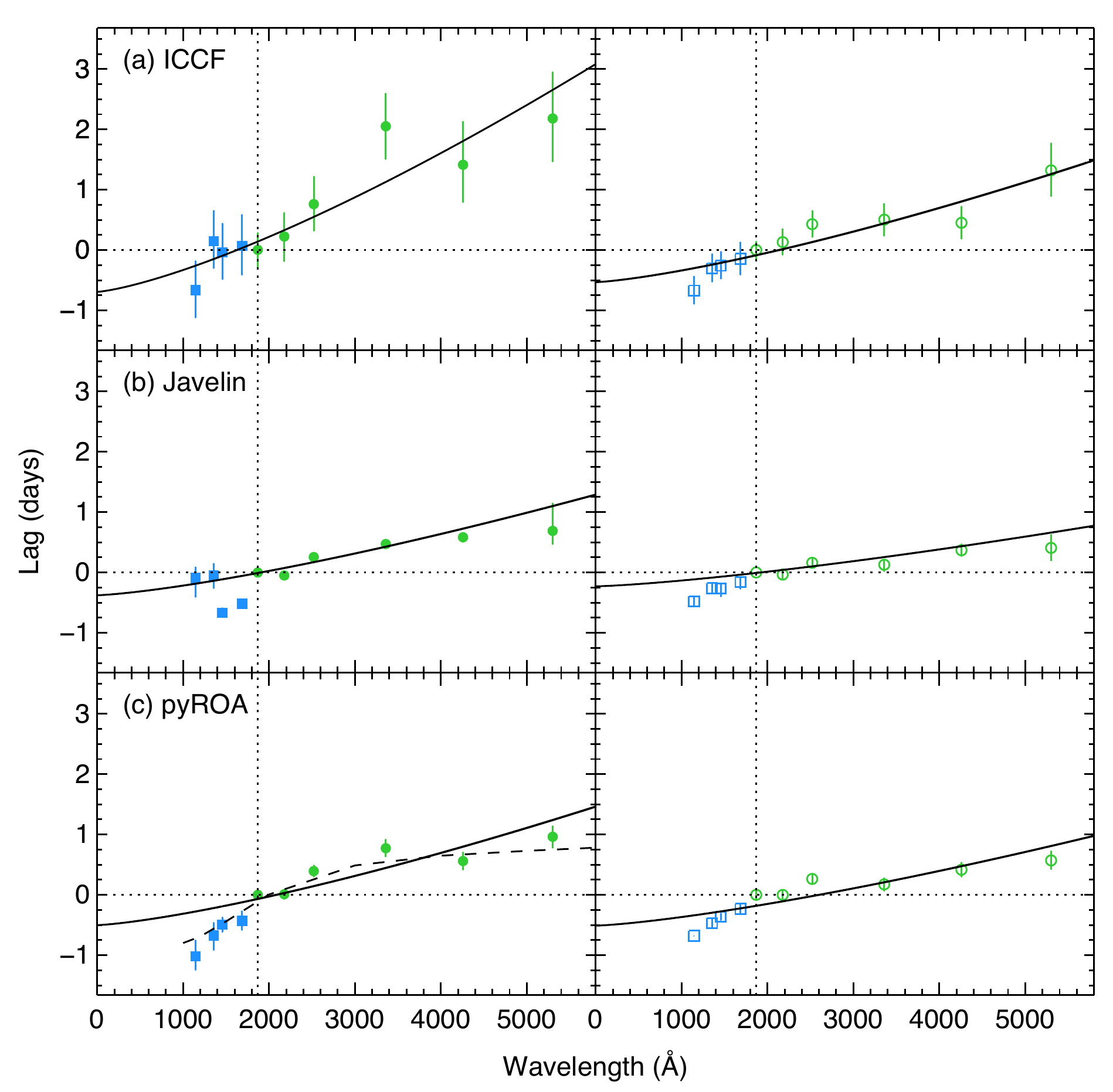}
\caption{Lags as a function of rest wavelength with respect to the {\it UVW2} band (vertical dotted line) for \swift\ (green) and \hst\ (blue) wave bands for the three different lag determination methods.  Left-hand panels (filled symbols) indicate lags calculated from the original light curves, while right-hand panels (open symbols) are lags determined from the detrended light curves. The solid lines show the best-fitting $\tau \propto \lambda^{4/3}$ relation to the lags using the original and detrended light curves.  The dashed line in panel (c) shows the best-fitting lag relation from the `bowl' model.}
\label{fig:lags}
\end{figure*}

The right-hand panels of Fig.~\ref{fig:alllc} show the CCF for each band, while the histograms show the distribution of lag centroid values from the FR/RSS method. For the UV/optical bands, the lag centroid is shown as a function of wavelength in Fig.~\ref{fig:lags}.  The lags generally increase with wavelength, roughly following $\tau \propto \lambda^{4/3}$, with a potential excess lag in the $U$-band and some scatter.  For all the \swift\ bands, the lag peaks are shorter than the lag centroids, though both still generally increase with increasing wavelength.  That the CCF peak lag is shorter than the centroid lag indicates that the CCF is asymmetric, which would indicate an asymmetric transfer function (since the CCF is the convolution of the auto-correlation function with the transfer function).  Disk transfer functions are asymmetric  \citep[e.g.,][]{cackett07, starkey16}, but this could also indicate emission from an extended region \citep{cackett22}.  The difference between peak and centroid is particularly noticeable in the $U$ band where the BLR diffuse continuum emission peaks locally.

As discussed earlier, the shortest wavelength light curves show a different long-term trend than the rest of the light curves.  We therefore also determine the lags from the detrended light curves.  We find this shortens the lags quite dramatically in most bands (e.g., more than a factor of 2 in $U$ and $B$), indicating that there is a contribution to the lags on long timescales.  The resulting lags are given in Tab.~\ref{tab:lags} and shown as open symbols in Fig.~\ref{fig:lags}.  The detrended centroid lags are consistent with the peak lags from the non-detrended light curves.  The $U$-band excess is not observed in the detrended lags, suggesting the excess lag is caused by the variability on long timescales that is removed by detrending.

A number of previous studies show that lags can be timescale-dependent, with long-term variations significantly impacting the measured lag \citep[e.g.,][]{mchardy14,mchardy18,pahari20,vincentelli21}.  Moreover, recent work investigating frequency-dependent time lags in NGC~5548 showed that it was low-frequency (long-timescale) variations from large sizescales that dominate the CCF measurement \citep{cackett22}. The difference between the lags that we measure in \mrk\ once detrended also supports this.
 These long-term trends are thought to be due to variability from more distant regions, or some other source of variability in the disk, and so need to be removed in order to isolate the short-term variations expected to be from inner disk reverberation.  We searched for lags between the detrending functions themselves, but did not measure anything significant.
 
We also try calculating the lags by using the non-detrended light curves only after day 9330, ignoring the early part of the light curve where the scaling discrepancy occurs.  The lag centroids in all bands are consistent within 1$\sigma$ with lags from the full light curve.  However, the $U$, $B$, and $V$-band lags all drop by approximately 0.9 days.  We leave a more detailed investigation of lags during different periods of the light curves to future work.

\subsubsection{\sc Javelin}

The {\sc Javelin} analysis package \citep{zu11,zu13} uses the Markov Chain Monte Carlo (MCMC) method to model the variability of the reference light curve assuming a damped random walk to model the light curve.  {\sc Javelin} is particularly useful when there are gaps in the light curves, since it uses information on the variability properties of the light curve to interpolate between the gaps.  {\sc Javelin} has been shown to produce estimates of the uncertainties that are closer to the input of simulations than the ICCF method \citep{yu20}.  

To measure lags between different bands, one first fits the reference band light curve with a damped random walk model.  {\sc Javelin} then takes the posteriors on the variability parameters from this fit and shifts, smooths and scales the light curve to fit the other band.  The shifting and smoothing assumes a top-hat transfer function.  For all bands with wavelengths longer than {\it UVW2}, we use the {\it UVW2} as the reference light curve.  For the bands shorter than {\it UVW2} we use the band of interest as the reference, and {\it UVW2} as the responding light curve, and then flip the sign of the lags.  We fit pairs of light curves (wave band of interest plus reference band) rather than fitting all light curves simultaneously (with a different lag for each light curve) as we found it did not easily converge with so many light curves at one time.

\begin{figure*}
\centering
\includegraphics[width=0.95\textwidth]{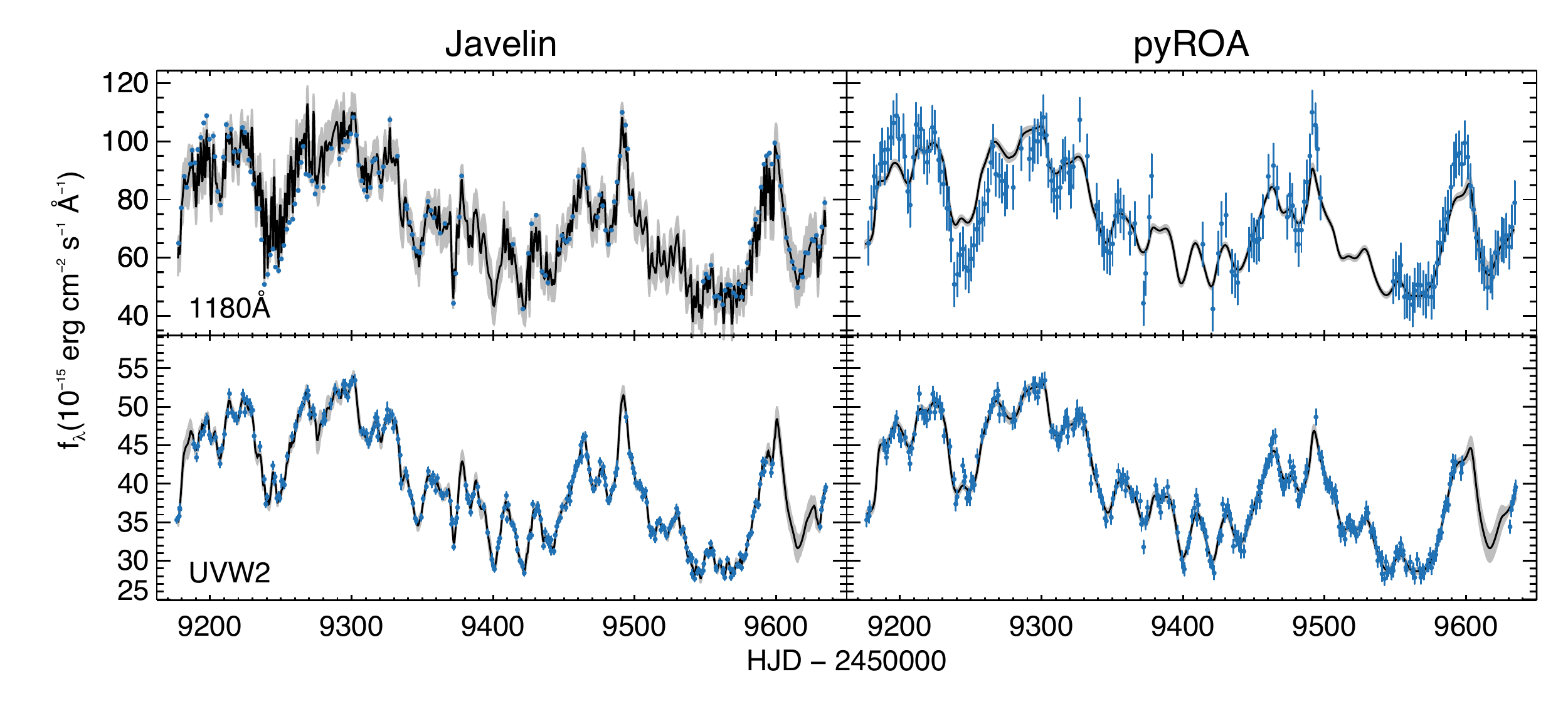}
\caption{Fits to the 1180~\AA\ (top) and {\it UVW2} (bottom) light curves found using {\sc Javelin} (left) and {\sc pyROA} (right).  {\sc Javelin} can only achieve a good fit by introducing unrealistic large-amplitude, short-timescale variability.  {\sc pyROA} achieves a good fit by significantly inflating the 1180~\AA\ error bars, but the model still does not match all the peaks and troughs in the light curves. The failure of both methods is a consequence of analyzing light curves that are not simply shifted, smoothed, and scaled versions of each other.}
\label{fig:javpyroa}
\end{figure*}

\begin{figure*}
\centering
\includegraphics[width=0.95\textwidth]{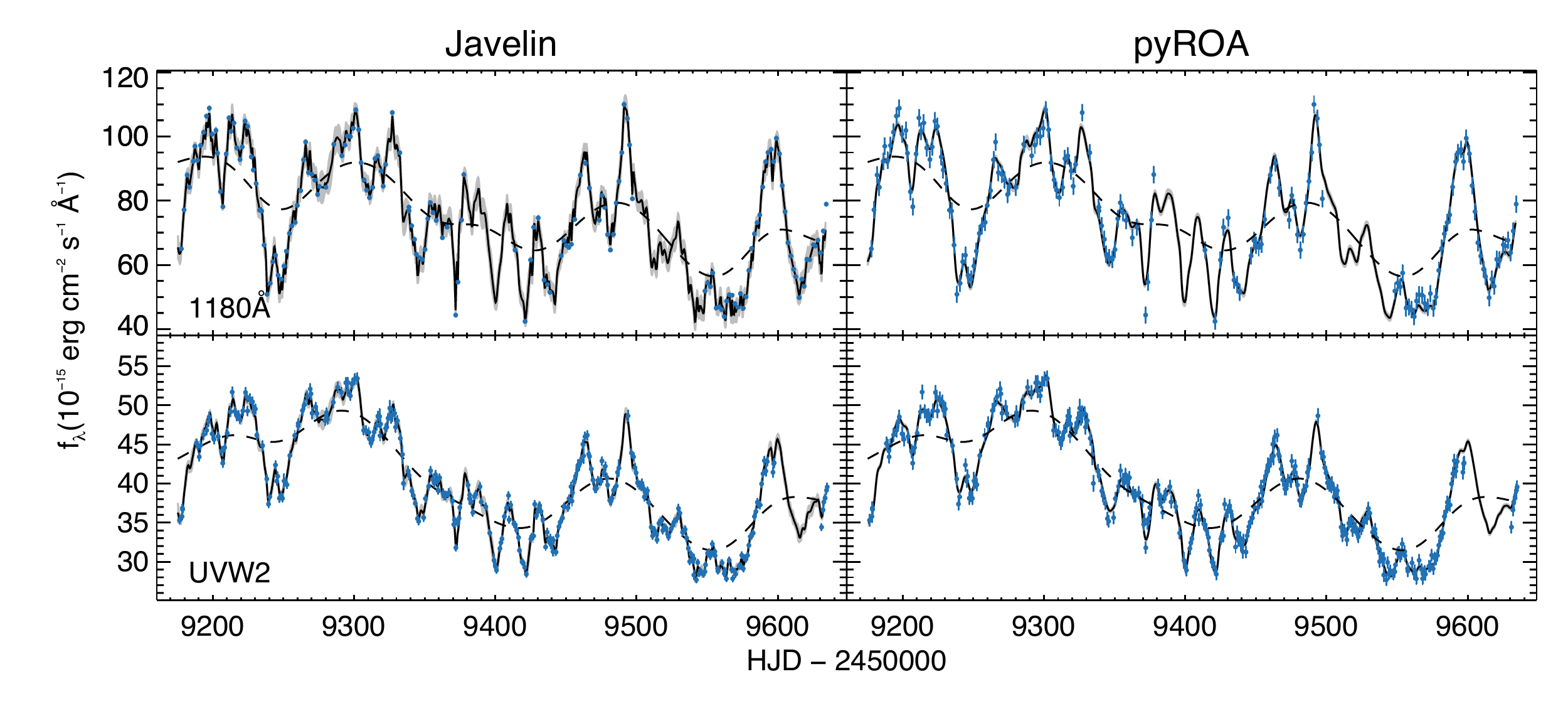}
\caption{Fits to the detrended 1180~\AA\ (top) and {\it UVW2} (bottom) light curves using {\sc Javelin} (left) and {\sc pyROA} (right).  The dashed line shows the detrending light curve.  Significantly improved fits are achieved with both {\sc Javelin} and {\sc pyROA} once the light curves are detrended.}
\label{fig:javpyroa_detrend}
\end{figure*}

We find that when modelling the {\it UVW2} and \hst\ light curves the model overfits the data, with unrealistic variability on short timescales to try to match every variation in the light curve (see Fig.~\ref{fig:javpyroa}).  This is indicative of the different long-term trends in the light curves -- they break the assumption that the light curves are simply shifted, smoothed, and scaled versions of each other.  Interestingly, if we omit the first portion of the light curve (before day 9330) then {\sc Javelin} can fit the light curves without this problem. We also find that when modeling the detrended light curves we do not encounter this problem either, and find a good fit to the light curves and well-recovered lags in all bands (see Fig.~\ref{fig:javpyroa_detrend}).  {\sc Javelin} lags are given in Table~\ref{tab:lags} and shown in panel (b) of Fig.~\ref{fig:lags}.  The {\sc Javelin} lags are much closer to the ICCF peak lags than the centroid lags.

\subsubsection{PyROA}

The {\sc pyROA} analysis package \citep{donnan21} takes a Bayesian MCMC approach to fitting the light curves.  The light curves are modeled using a running optimal average -- each point in the light curve is an inverse-variance weighted average of the data within a window function. The window function reduces the weight of the data points far from the time of interest (see equation 1 in \citealt{donnan21}).  The algorithm fits the model to determine the mean and root-mean square (rms) of each light curve as well as the width of the running optimal average window function and the lag between each of the light curves.  Moreover, {\sc PyROA} fits for additional variance to expand the uncertainties in each light curve, this accounts for cases where uncertainties in the data have been underestimated.  In principle, it can be run to fit all light curves together, but, as described above, the differing long-term trends between the \hst\ and the {\it UVW2} band prevent achieving a good fit.  We therefore fit each of the light curves individually, but find issues in those shortest bands -- to achieve a reasonable fit {\sc pyROA} must expand the uncertainties significantly, and still the best-fit model underpredicts some peaks and troughs (see Fig.~\ref{fig:javpyroa}).  Similar to what we found with {\sc Javelin}, the varying long-term trend breaks the assumption of the light curves simply being shifted, smoothed, and scaled versions of each other.

\input{tab2.tex}

{\sc pyROA} has the option to fit a variable background to each light curve (detrending). However, to be consistent with the other analysis techniques, we fit the detrended light curves.  We find all detrended light curves can be fit together, resulting in good fits to the light curve (see Fig.~\ref{fig:javpyroa_detrend}) and well-determined lags in each band.  This method requires the shortest wavelength (1180~\AA) light curve to be used as the reference and assumed to have zero lag, which differs from the other approaches that have used {\it UVW2} as the reference.  In Table~\ref{tab:lags} we give the {\sc pyROA} lags, and they are also shown in panel (c) of Fig.~\ref{fig:lags}.  For the lags from fits to the detrended light curves, we shift all lags by the 1180~\AA\ to {\it UVW2} lag for direct comparison to lags determined with other methods.

\subsection{Spectral Analysis}

\begin{figure}
\centering
\includegraphics[width=\columnwidth]{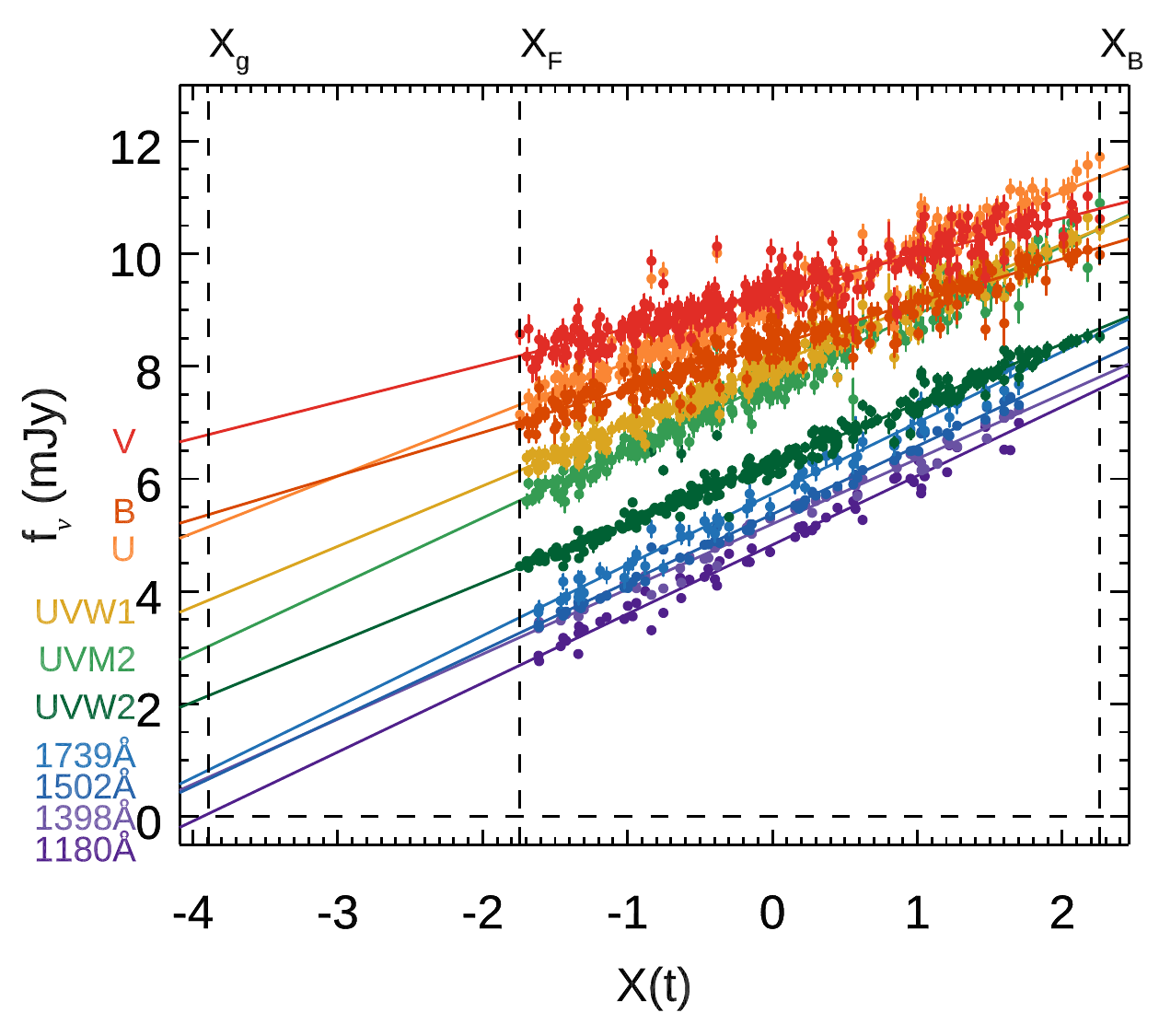}
\caption{Flux vs. the model light curve $X(t)$ for all wavebands.  The best-fitting flux-flux relations are shown as solid lines.  $X_g$ indicates the value of $X(t)$ where the error envelope of the fit to the shortest wavelength band (1180~\AA; purple) intercepts $f_\nu$ = 0.  The flux-flux relations at $X(t) = X_g$ gives the value of the constant flux in each band, while the slope of the relations give the strength of the variable component.  $X_F$ and $X_B$ indicate the faint and bright values of $X(t)$.}
\label{fig:Xtfl}
\end{figure}

We can use the light curves to determine the spectrum of the variable and constant components using the flux-flux analysis method \citep[also sometimes referred to as the flux variation gradient method, see, e.g.,][for recent examples]{mchardy18,hernandezsantisteban20,cackett20,fian22}.  We deredden the observed light curves assuming Galactic absorption of $E(B-V) = 0.022$ mag and a \citet{cardelli89} extinction law with $R_V = 3.1$.  The flux densities are also corrected to the rest frame (through $f_\lambda / (1 + z)^3$). We fit the light curves with a linear model using a dimensionless light curve $X(t)$ that is normalized to have a mean of 0 and standard deviation of 1.  The model light curve for each band is then a constant, $A_\nu(\lambda)$, plus $X(t)$ multiplied by a scale factor $S_\nu(\lambda)$, 
\begin{equation}
f_\nu(\lambda, t) = A_\nu(\lambda) + S_\nu(\lambda)X(t) \; . 
\end{equation}
In other words, the variability in each band is a shifted and scaled version of the dimensionless light curve $X(t)$, and there is no change in the shape of the spectral energy distribution (SED) of the variable component.  The light curves are fit simultaneously with $X(t)$, $A_\nu(\lambda)$, and $S_\nu(\lambda)$ as free parameters.  Since the \hst\ and \swift\ data are not sampled at the same cadence, we only include data points within 0.5 days of the {\it UVW2} observations.  We do not correct for time lags, though lags should only add scatter to the flux-flux plot.  The scale factor $S_\nu(\lambda)$ gives the rms spectrum of the variable component of the light curves.  We determine the constant component of the spectrum, $A_\nu(\lambda)$, by evaluating the best-fitting flux-flux relations at $X(t) = X_g$ defined by where the error envelope of the fit to the 1180~\AA\ band intercepts $f_\nu$ = 0 (see Fig.~\ref{fig:Xtfl}). The maximum and minimum spectral energy distributions are determined at $X(t) = X_B$ and $X_F$ respectively.

\begin{figure}
\centering
\includegraphics[width=\columnwidth]{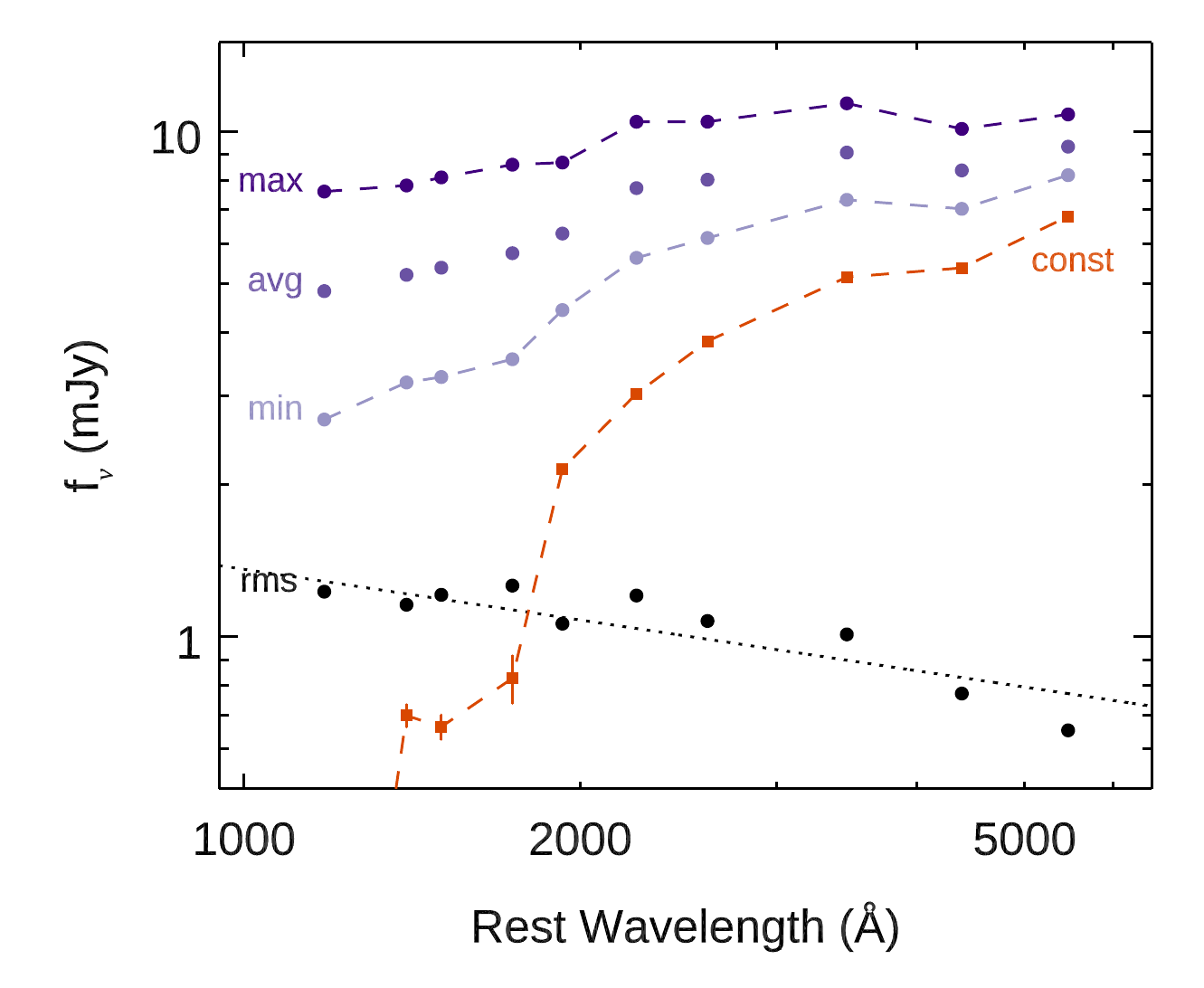}
\caption{The UV/optical SED of \mrk\ during AGN STORM 2.  The variable (rms, black circles) spectrum approximately follows $f_\nu \propto \lambda^{-1/3}$ as expected for an accretion disk.  The constant component is shown as orange squares. The purple circles show the maximum, mean and minimum spectral energy distribution.  Error bars are plotted, but are mostly smaller than the symbols.}
\label{fig:flflsed}
\end{figure}

We show the flux-flux relations in Fig.~\ref{fig:Xtfl}, and the resulting SED of \mrk\ in Fig.~\ref{fig:flflsed}.  The best-fitting parameters are given in Tab.~\ref{tab:flfl}.  The rms spectrum approximately follows the relation $f_\nu \propto \lambda^{-1/3}$ expected for an accretion disk, though more realistic disk models \citep[e.g.,][]{slone12} give a spectrum that is closer to constant in $f_\nu$ over the observed wavelength range for the black hole mass and mass accretion rate of Mrk 817. The constant component shows a strong increase at around 2000~\AA\ that may be associated with a strong non-variable, or slowly varying, \ion{Fe}{2} component and/or diffuse continuum.  The host-galaxy flux below 3000~\AA\ should be minimal compared to the AGN, while at longer wavelengths the host-galaxy contribution becomes important.  That the variable SED follows $f_\nu \propto \lambda^{-1/3}$ and does not turn over significantly at the blue end suggests that intrinsic reddening of the continuum in \mrk\ is not significant.  This is in contrast to \citet{jaffarian20} who determine intrinsic reddening of $E(B-V) = 0.55$ for \mrk.  \citet{merritt22} also finds a lower intrinsic reddening of $E(B-V) = 0.1$ to 0.2.

\input{tab3.tex}
\section{Discussion and Summary}\label{sec:discuss}
We monitored \mrk\ for approximately 15 months with \swift\ in the X-ray and six UV/optical bands as part of the AGN STORM 2 campaign.  The 0.3 -- 10 keV X-ray count rates are on average a factor of 6 fainter than archival observations, and show suppressed variability aside from a large flare (factor of 10 increase) which peaks close to the historical mean flux.  Analysis of the X-ray hardness-intensity diagram shows that the \swift\ X-ray spectra are consistent with being heavily obscured, with variability in the absorption throughout.  This matches what is seen in higher quality XMM and NuSTAR spectra \citep{kara21, miller21}.  Tracking of the variable X-ray spectrum with NICER is explored in more detail in Paper III \citep{partington23}.

The \swift\ UV/optical light curves are highly variable throughout the campaign.  Despite the X-ray count rate being significantly fainter, the UV flux remains almost unchanged when compared to archival data.  The X-ray band is poorly correlated with the UV/optical light curves throughout, with a maximum correlation coefficient of $R_{\rm max} = 0.33$~days.  This was also seen during the previous \swift\ monitoring of \mrk\  \citep{morales19}.  During that previous campaign, X-ray obscuration does not appear to have been significant, while our current campaign had significant obscuration throughout.  The lack of correlation may be explained if the obscuration seen by the continuum-emitting region is not what we see along our line of sight, or if X-ray variations are not significantly affecting the emitted optical/UV flux or vice versa.

The UV/optical light curves, both HST and Swift, are well-correlated with each other throughout.  We measure interband continuum lags that generally increase with increasing wavelength. Interestingly, we find that the light curves show a period at the beginning where the strength of the variations in the continuum is suppressed compared to later periods -- the shortest wavelength light curve (1180~\AA) cannot be simply shifted and scaled to match the longer wavelength light curves with one scale factor for the full campaign (see Fig.~\ref{fig:1180W2}).   While the slope of the relation between the two bands is approximately the same throughout, the normalization of the relation changes.  This discrepant period can also be seen by comparing scaled versions of the light curves directly (Fig.~\ref{fig:W2comp}).  There, the significant double trough towards the beginning of the light curve is significantly less deep in the longer wavelength bands compared to 1180~\AA, while variations in the latter part of the light curve are well matched.  

\begin{figure}
\centering
\includegraphics[width=\columnwidth]{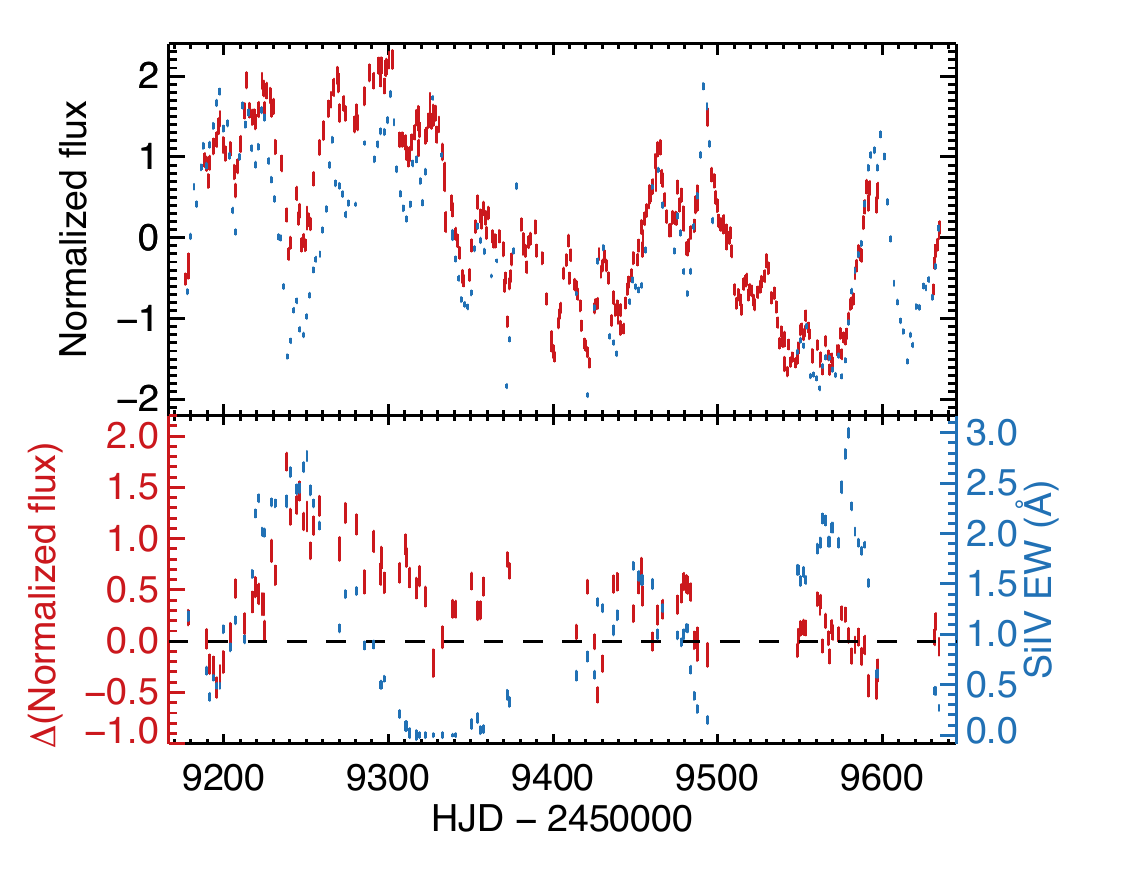}
\caption{{\it Top:} The normalized {\it UVW2} (red) and 1180~\AA\ (blue) light curves. {\it Bottom:} The difference between the normalized {\it UVW2} and 1180~\AA\ fluxes (red) as compared to the strength of the UV absorption, as measured by the EW of the broad \ion{Si}{4} absorption trough (blue).}
\label{fig:deltaf_vs_ew}
\end{figure}

This double trough occurs at the time in the first half of the light curve when the UV and X-ray absorption is strongest, as measured by the equivalent width of the broad \ion{Si}{4} absorption trough and the X-ray, $N_{\rm H}$, column density \citep{kara21}.  We demonstrate this in Fig.~\ref{fig:deltaf_vs_ew} by comparing the difference in flux between the normalized {\it UVW2} and 1180~\AA\  light curves with the evolution of the broad \ion{Si}{4} absorption equivalent width (EW), calculated from the \hst\ spectra following \citet{kara21} and to be presented in detail in a future paper.  In the early part of the light curve, the growth of the discrepancy between the light curves follows the increase in absorption, and the reduction in the discrepancy follows the decrease in absorption.  The discrepancy has a maximum approximately when the absorber is strongest, and the light curves come into agreement again shortly after the absorption reaches a minimum.  The light curves remain similar after this, despite a large increase in absorption toward the very end of the campaign.  

It is tempting to associate the discrepancy in fluxes at the beginning of the campaign with the presence of increased absorption.  It is interesting that this early part of the light curve is also where the lag of the \ion{C}{4} emission line changes dramatically, dropping from around 12 days in the first time segment to 2 days in the second (Homayouni et al., submitted).  As discussed in that paper, this anomalous behavior is potentially explained by the high obscuration early on -- the appearance of an obscuring screen between the ionizing radiation and the BLR will cause the nearby material to stop responding first, giving rise to a long lag.  The disappearance of this screen will weight the response to smaller radii again.  Moreover, radiation absorbed by the obscurer must be reemitted, and \citet{dehghanian19b} predict that this reemitted flux is largely in enhanced diffuse continuum, and in the broad wings of the emission lines. This enhanced diffuse continuum emission could be appearing in the Swift UV bands. That we see anomalous behavior in the continuum at the same time as this anomalous behavior in the emission lines and the high obscuration potentially suggests a significant fraction of the continuum emission arises in the BLR at all wavelengths.  This is similar to behavior in NGC~5548 where during the anomalous `BLR holiday' when emission-line variations decorrelated from the continuum \citep{goad16}, the continuum bands also showed a change in correlation \citep{goad19}.  This was used to argue a significant fraction of the continuum arises from the BLR, not the disk.  Here too, the behavior we observe suggests a significant fraction of the continuum arises from the BLR, though careful modeling \citep[e.g.,][]{koristagoad19,netzer22} is needed to determine exactly how much. 

Alternatively, it could simply be that there is additional underlying disk variability that is not associated with reverberation.  Variability on long timescales (thousands of days) is usually not associated with reverberation \citep[e.g.,][]{breedt09}, but in this case the anomalous period is much shorter, just 100 or so days.  However, recent work to create temperature maps from AGN light curves \citep{neustadt22} shows temperature fluctuations on timescales of tens to hundreds of days that are not associated with reverberation. What we are seeing here in Mrk~817 could also be temperature fluctuations in the disk not associated with reverberation.

Since the continuum variability breaks one of the fundamental assumptions of light curve modeling methods (that the light curves are blurred, shifted, and scaled versions of one another), such methods ({\sc Javelin} and {\sc pyROA}) do not do a good job fitting the original light curves.  We overcome this by detrending the light curves to account for the suppressed response at the beginning of the campaign.  This allows for good fits with both {\sc Javelin} and {\sc pyROA} (compare Fig.~\ref{fig:javpyroa} and Fig.~\ref{fig:javpyroa_detrend} for fits without and with detrending), and well-determined lags using these methods.  The {\sc Javelin} and {\sc pyROA} lags are significantly shorter than the lag centroid determined using the ICCF method on the unadjusted light curves, and closer to the lag peak.  The ICCF lag centroid from the detrended light curves more closely matches the {\sc Javelin} and {\sc pyROA} lags.  The detrending we perform is higher-order than the linear or low-order polynomial more commonly used, but is required given the relatively sharp and short-term discrepancy in the continuum bands. While this allows for successful modeling of the light curves it may be removing some of the real lag signal on long timescales.

\begin{figure}
\centering
\includegraphics[width=\columnwidth]{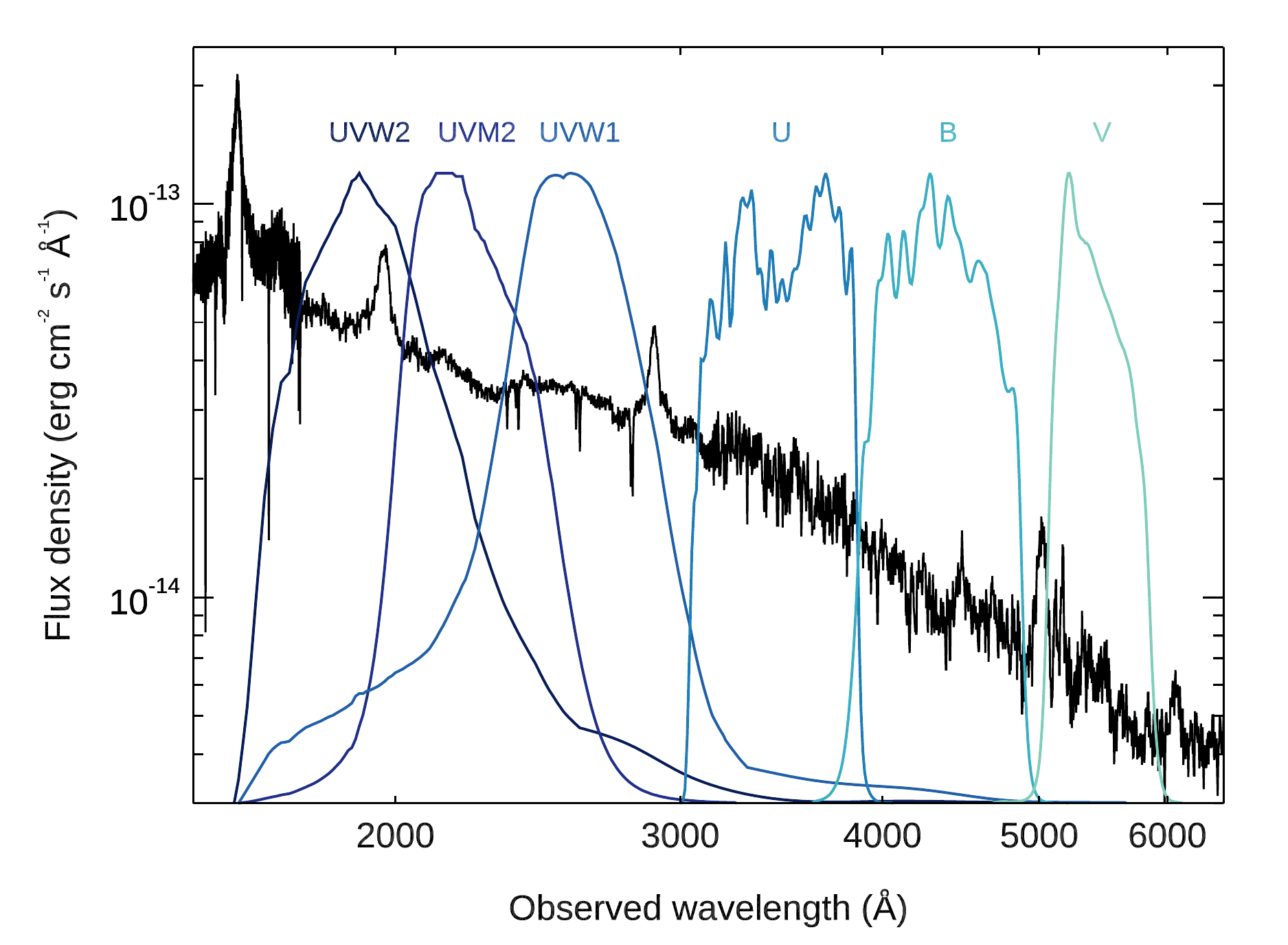}
\caption{HST/STIS spectrum of Mrk 817 from 2021 April 18 compared to the Swift filter transmission curves.  Several strong broad emission lines fall within the Swift filters.}
\label{fig:hstspec}
\end{figure}

The difference between the lag centroid and lag peak from the unadjusted light curves indicates an asymmetric transfer function -- that there is significant response on longer timescales.  This is supported by the decrease in lag centroid once the light curves are detrended.  An excess lag in the $U$ band is seen in analysis of the unadjusted light curves in the lag centroids using the ICCF technique and in {\sc pyROA} fits.  This $U$-band excess is almost universally seen in other continuum reverberation-mapping and has been attributed to diffuse continuum emission from the BLR \citep{koristagoad01, koristagoad19,lawther18,netzer20,netzer22}.  While it peaks in the $U$ band at the Balmer jump, and also at the Paschen jump, the BLR diffuse continuum should affect all wavebands.  Since the BLR emitting region is presumably more extended than the UV/optical part of the accretion disk, the timescale of the response from the BLR should be longer than from the accretion disk.  It is interesting, that when long-timescale variations are removed the $U$-band excess disappears.  This would again support the idea that significant continuum emission originates from the BLR.  Of course, the Swift filters are broadband and so include emission lines too.  Fig.~\ref{fig:hstspec} shows a broadband HST/STIS spectrum from our campaign compared to the Swift filter bandpasses.  Prominent emission lines \ion{C}{3}] $\lambda$1909, \ion{Mg}{2} $\lambda$2800 and H$\beta$ fall within the {\it UVW2}, {\it UVW1}, and {\it B} filters respectively. This will lengthen the lags in those bands, though simulations during other campaigns show this does not dominate \citep[e.g.,][]{fausnaugh16}.  See a similar discussion for the case of Fairall 9 in Edelson et al. (in prep.).  On the other hand, the HST continuum light curves are calculated over line-free integration windows \citep{homayouni23}.  A full analysis of the lags using frequency-dependent methods \citep{cackett22} and power spectral analysis is left to future work, as is a detailed modeling of the spectra to determine the strength of the expected BLR continuum lags following the methods of \cite{koristagoad19} and \cite{netzer22}.

We use a flux-flux analysis to separate the variable and constant components of the light curves.  The variable spectrum approximately follows $f_\nu \propto \lambda^{-1/3}$ as expected if the variable component is dominated by a geometrically thin, optically thick accretion disk.  The constant component shows a strong increase at  $\sim$2000~\AA\ that may be associated with strong non-variable (or slowly varying) \ion{Fe}{2} component and/or diffuse continuum.  At longer wavelengths the constant component can be attributed to the host-galaxy flux.

As noted in the Introduction, many previous continuum reverberation campaigns have found lags to be a factor of a few longer than expected given reasonable estimates for the mass and mass accretion rate.  Removing long-timescale variations shortens the lags, and these shorter lags are generally consistent with the expected disk size.  Equation 12 from \citet{fausnaugh16} gives the expected lag at the reference band, $\alpha$.  We use {\it UVW2} as the reference band, and assume $M = 3.85\times10^7$~M$_\odot$ and $\dot{m}_{\rm E} = 0.2$ for Mrk~817, along with assuming an accretion efficient of $\eta = 0.1$ and a local ratio of external to internal heating of $\kappa = 1$. This gives $\alpha = 0.31$ days.  Our best-fitting $\tau \propto \lambda^{4/3}$ relations to the detrended lags range from $\alpha = 0.23\pm0.04$ ({\sc Javelin}) to $\alpha = 0.53\pm0.13$ (ICCF), bracketing the expected disk size.  Thus, the lags on short timescales are broadly consistent with reverberation from a standard disk.

Some alternative models for continuum reverberation lags invoke different geometries.  One example is the model of \citet{starkey23}, where there is a steep rim occurring near the disk's dust sublimation radius (the `bowl' model).  We fit this model jointly to the bright and faint disk SED (determined from the flux-flux analysis) and the {\sc pyROA} lags (not detrended).  This model fits the SED and lags well.  The best-fitting lag relation is shown as a dashed line in panel (c) of Fig.~\ref{fig:lags}.  Note how the lag relation flattens off at longer wavelengths.  The inclusion of ground-based optical light curves at longer wavelengths will allow for a better test of this model.

Future AGN STORM 2 papers will present photometric and spectroscopic ground-based optical and near-IR data to probe more distant material, and more sensitive X-ray spectra from NuSTAR, and XMM to better trace the X-ray obscurer.  Extended monitoring of Mrk~817 with \swift, NICER, and ground-based monitoring continues and promises to offer further insights into this complex AGN. 

\begin{acknowledgements}
Our project began with the successful Cycle 28 HST proposal 16196 \citep{peterson20}.  E.M.C. gratefully acknowledges support from NASA through grant 80NSSC22K0089. 
Y.H. acknowledges support from the Hubble Space Telescope program GO-16196, provided by NASA through a grant from the Space Telescope Science Institute, which is operated by the Association of Universities for Research in Astronomy, Inc., under NASA contract NAS5-26555. Research at UC Irvine was supported by NSF grant AST-1907290. 
J.G. gratefully acknowledges support from NASA through grant 80NSSC22K1492.
CSK is supported by NSF grant AST-2307385.
MCB gratefully acknowledges support from the NSF through grant AST-2009230. 
H.L. acknowledges a Daphne Jackson Fellowship sponsored by the Science and Technology Facilities Council (STFC), UK.   
D.I., A.B.K, and L.\v C.P. acknowledge funding provided by the University of Belgrade - Faculty of Mathematics (the contract 451-03-68/2022-14/200104), Astronomical Observatory Belgrade (the contract 451-03-68/2022-14/ 200002), through the grants by the Ministry of Education, Science, and Technological Development of the Republic of Serbia. 
D.I. acknowledges the support of the Alexander von Humboldt Foundation. 
A.B.K. and L.{\v C}.P thank the support by Chinese Academy of Sciences President's International Fellowship Initiative (PIFI) for visiting scientist. 
A.V.F. is grateful for financial assistance from the Christopher R. Redlich Fund and numerous individual donors. 
MV gratefully acknowledges support from the Independent Research Fund Denmark via grant number DFF 8021-00130. 
Y.R.L. acknowledges financial support from NSFC through grant Nos. 11922304 and 12273041 and from the Youth Innovation Promotion Association CAS. 
This work made use of data supplied by the UK Swift Science Data Centre at the University of Leicester.
\end{acknowledgements}

\facilities{Swift, HST (COS)} 

\software{{\sc HEAsoft} \citep{heasoft}, {\sc xspec} \citep{arnaud96}, {\sc Javelin} \citep{zu11,zu13}, {\sc pyROA}  \citep{donnan21} }

\bibliographystyle{aasjournal}
\bibliography{agn}

\end{document}

%% file: tab2.tex
\movetabledown=5cm
\begin{rotatetable*}
\begin{deluxetable*}{lcCCCCCCCCCc}
\label{tab:lags}
\tablecaption{Time lags (rest frame) calculated with respect to {\it UVW2}, along with the variability amplitude and maximum correlation coefficient.}
\tabletypesize{\scriptsize}
\tablewidth{0pt}
\tablehead{
 \colhead{Filter} & $\lambda_{\rm cent}$ & \colhead{$\tau_{\rm cent}~{\rm (days)}$} & \colhead{$\tau_{\rm peak}~{\rm (days)}$} & \colhead{$\tau_{\rm cent}~{\rm (days)}$} & \colhead{$\tau_{\rm peak}~{\rm (days)}$} & \colhead{$\tau~{\rm (days)}$} & \colhead{$\tau~{\rm (days)}$} & \colhead{$\tau~{\rm (days)}$} & \colhead{$\tau~{\rm (days)}$} & \colhead{$F_{\rm var}$} & \colhead{$R_{\rm max}$} \\
  & (observed) & {\rm ICCF} & {\rm ICCF} & {\rm Detrend} & {\rm Detrend} & \textsc{Javelin}& \textsc{Javelin}, {\rm Detrend} &  \textsc{pyROA} & \textsc{pyROA}, {\rm Detrend} }
\decimalcolnumbers
\startdata
 X-ray & 0.3 -- 10 keV & - & - & - & - & - & - & - & - & 1.071\pm0.004 & 0.33 \\
 HST/COS & 1180\,\AA      & -0.66_{-0.46}^{_+0.48} & -0.68_{-0.19}^{+0.24} & -0.67\pm0.24 & -0.58\pm0.19          & -0.09_{-0.32}^{+0.18} & -0.47\pm0.09          & -1.02_{-0.23}^{+0.27}  & -0.68 &  0.225\pm0.001 &  0.88 \\
 HST/COS & 1398\,\AA      & 0.15_{-0.45}^{+0.51}   & -0.34\pm0.19          & -0.30\pm0.24 & -0.39_{-0.15}^{+0.19} & -0.05\pm0.21          & -0.26\pm0.07          & -0.67_{-0.24}^{+0.22}  & -0.47\pm0.07 &  0.204\pm0.001 & 0.90 \\
 HST/COS & 1502\,\AA      & -0.04_{-0.45}^{+0.48}  & -0.34\pm0.19          & -0.26\pm0.23 & -0.34_{-0.15}^{+0.19} & -0.67_{-0.05}^{+0.08} & -0.27_{-0.14}^{+0.07} & -0.50\pm0.13           & -0.36\pm0.06 &  0.202\pm0.001 & 0.95 \\
 HST/COS & 1739\,\AA      & 0.07_{-0.49}^{+0.52}   & -0.24_{-0.24}^{+0.34} & -0.15\pm0.28 & -0.24\pm0.24          & -0.52\pm0.07          & -0.16_{-0.12}^{+0.09} & -0.43\pm0.16           & -0.23\pm0.08 &  0.192\pm0.002 & 0.94 \\
 {\it UVW2} & 1928\,\AA   & 0.00\pm0.29 	   & 0.00\pm0.10           & 0.00\pm0.15  & 0.00\pm0.10           & 0.00\pm0.01           & 0.00\pm0.01           & 0.00\pm0.05            & 0.00\pm0.08 &  0.169\pm0.001 & 1.00\\
 {\it UVM2} & 2246\,\AA   & 0.23\pm0.40 	   & -0.19_{-0.15}^{+0.53} & 0.13\pm0.22  & -0.19_{-0.15}^{+0.49} & -0.05\pm0.08          & -0.03\pm0.07          & 0.01\pm0.08            & 0.00\pm0.08 &  0.156\pm0.001 & 0.99\\
 {\it UVW1} & 2600\,\AA   & 0.76\pm0.46 	   & 0.39_{-0.05}^{+0.10}  & 0.43\pm0.22  & 0.34\pm0.10           & 0.25\pm0.07           & 0.16\pm0.07           & 0.39\pm0.10            & 0.26\pm0.09 &  0.138\pm0.001 & 0.98 \\
 $U$ & 3465\,\AA          & 2.05\pm0.55 	   & 0.48_{-0.15}^{+0.10}  & 0.50\pm0.27  & 0.29_{-0.53}^{+0.15}  & 0.47\pm0.07           & 0.13_{-0.11}^{+0.09}  & 0.77\pm0.15            & 0.17\pm0.11 &  0.115\pm0.001 & 0.97 \\
 $B$ & 4392\,\AA          & 1.41_{-0.62}^{+0.72}   & 0.48_{-0.19}^{+0.29}  & 0.45\pm0.27  & 0.39_{-0.15}^{+0.19}  & 0.58\pm0.08           & 0.37\pm0.11           & 0.56\pm0.15            & 0.42\pm0.12 &  0.094\pm0.001 & 0.95 \\
 $V$ & 5468\,\AA          & 2.18_{-0.72}^{+0.77}   & 0.68_{-0.34}^{+1.36}  & 1.32\pm0.45  & 0.53_{-0.39}^{+1.07}  & 0.69_{-0.22}^{+0.46}  & 0.41\pm0.22           & 0.96\pm0.19            & 0.57\pm0.16 &  0.071\pm0.001 & 0.94 \\
\enddata
\tablecomments{The lags measured using \textsc{pyROA} fits to the detrended light curves were measured relative to the shortest wavelength (1180~\AA) light curve and fit simultaneously. The 1180~\AA\ band therefore has a lag fixed to zero, but,  to directly compare to other methods measuring lags relative to {\it UVW2} we give the lags here offset by the 1180~\AA\ to {\it UVW2} lag. Swift filter central wavelengths are from \citet{poole08}.}
\end{deluxetable*}
\end{rotatetable*}

%% file: tab3.tex
\begin{deluxetable*}{lCCCCC}
\label{tab:flfl}
\tablecaption{Best-fitting parameters from the flux-flux analysis}
\tabletypesize{\scriptsize}
\tablewidth{0pt}
\tablehead{
 \colhead{Waveband} & \colhead{Avg. flux} & \colhead{Max. flux} & \colhead{Min. flux} & \colhead{$S_\nu$} & \colhead{$A_\nu$}}
\decimalcolnumbers
\startdata
1180\,\AA &
 4.828 \pm  0.012 &
 7.602 \pm  0.025 &
 2.690 \pm  0.021 &
 1.228 \pm  0.010 &
 0.051 \pm  0.041 \\
1398\,\AA & 
 5.199 \pm  0.011 &
 7.813 \pm  0.023 &
 3.185 \pm  0.019 &
 1.157 \pm  0.009 &
 0.698 \pm  0.037 \\
1502\,\AA & 
 5.373 \pm  0.011 &
 8.107 \pm  0.024 &
 3.265 \pm  0.020 &
 1.210 \pm  0.009 &
 0.663 \pm  0.038 \\
1739\,\AA & 
 5.739 \pm  0.024 &
 8.590 \pm  0.056 &
 3.540 \pm  0.046 &
 1.262 \pm  0.023 &
 0.827 \pm  0.091 \\
{\it UVW2}& 
 6.277 \pm  0.008 &
 8.674 \pm  0.015 &
 4.429 \pm  0.013 &
 1.061 \pm  0.006 &
 2.148 \pm  0.024 \\
{\it UVM2}& 
 7.720 \pm  0.013 &
10.446 \pm  0.028 &
 5.619 \pm  0.023 &
 1.206 \pm  0.011 &
 3.026 \pm  0.044 \\
{\it UVW1}& 
 8.022 \pm  0.011 &
10.449 \pm  0.024 &
 6.151 \pm  0.020 &
 1.074 \pm  0.010 &
 3.842 \pm  0.039 \\
$U$ & 
 9.079 \pm  0.011 &
11.362 \pm  0.025 &
 7.319 \pm  0.021 &
 1.010 \pm  0.010 &
 5.147 \pm  0.041 \\
$B$ &  
 8.369 \pm  0.010 &
10.113 \pm  0.024 &
 7.025 \pm  0.019 &
 0.772 \pm  0.010 &
 5.366 \pm  0.038 \\
$V$ &  
 9.326 \pm  0.011 &
10.801 \pm  0.027 &
 8.190 \pm  0.022 &
 0.653 \pm  0.011 &
 6.787 \pm  0.044 \\
\enddata
\tablecomments{All fluxes are $f_\nu$ in mJy and are rest-frame and extinction corrected.}
\end{deluxetable*}